\documentclass{article}

\usepackage{arxiv}

\usepackage[utf8]{inputenc} 
\usepackage[T1]{fontenc}    
\usepackage{hyperref}       
\usepackage{url}            
\usepackage{booktabs}       
\usepackage{amsfonts}       
\usepackage{nicefrac}       
\usepackage{microtype}      
\usepackage{lipsum}

\usepackage{graphicx}

\title{Architectural Software Patterns for the Development of IoT Smart Applications}

\author{
    Fabrizio F. Borelli \\
    Federal University of ABC (UFABC)\\
    Santo André- SP - BR \\
    \texttt{fabrizio.borelli@ufabc.edu.br} \\
   \And
    Gabriela Biondi \\
    Federal University of ABC (UFABC) \\
    Santo André- SP - BR \\
    \texttt{gabriela.biondi@ufabc.edu.br} \\
  \AND
    Fl\'avio Horita  \\
    Federal University of ABC (UFABC)\\
    Santo André- SP - BR \\
    \texttt{flavio.horita@ufabc.edu.br} \\
  \And
   Carlos Kamienski \\
   Federal University of ABC (UFABC) \\
   Santo André- SP - BR \\
   \texttt{carlos.kamienski@ufabc.edu.br} \\
}

\begin{document}
\maketitle

\begin{abstract}

Software developers usually start coding an application with no formal architecture in mind and relying on intuition and experience instead of on well-known design patters. A different approach is recommended for the development of IoT smart applications due to its high complexity that combines sensors, actuators, communication technologies, and big data analytics, as well as its distributed nature that spans for different layers of field, fog, and cloud infrastructure. Literature reports many experiences of software development for IoT smart applications. However, architectural solutions are presented with no rationale for the choice of software components and the way they relate to each other. This paper proposes a classification for software components and their relationships in order to model a software architecture for a particular IoT smart application. Three smart applications for cities, buildings, and agriculture were selected as examples of using some components, connectors, and well-known design patterns. Finally, the problems and challenges involved in the choice of software architectures for IoT are discussed.

\end{abstract}


\section{Introduction}
The increasing use of software has been allowing considerable changes in business management and the growing revenues of companies all over the world. Software architecture involves the structure and the organization of components and subsystems to interact among themselves, aiming at assembling systems \cite{Kruchten-2006}. Developers often start coding an application without a previously defined software architecture, that is, without having a sense of how several components and connectors will communicate to assemble a new system. Hence, many projects adopt the traditional layered software architecture, also known as n-tier architecture, which separates the source code into modules and packages \cite{shaw2006golden}.

Unfortunately, this practice often results in a disorganized collection of source code with modules that lack clear roles, responsibilities, and relationships \cite{Freeman-2004}. This poor software development practice generates solutions that are fragile, difficult to change, and costly to maintain \cite{Richards-2015}. 

The Internet of Things (IoT) brings additional complexity to software development due to its inherent distributed nature and the massive number of heterogeneous connected devices (sensors and actuators) \cite{Weyrich-2016}. The process of developing software architectures for IoT involves the interaction of a wide variety of components that play distinct roles. Although there are already initiatives to build these architectures, some challenges remain, such as\cite{Aksu-2018,Taivalsaari-2017}: 1) connectivity: things need to be connected to follow the IoT paradigm; 2) languages and tools: although there are several programming languages, there is no tool capable of orchestrating thousands of devices in a real-life scenario; 3) natural dynamics of IoT systems: the emergence of new communication protocols, new devices, and new programming languages renders the development of IoT solutions more complex \cite{Taivalsaari-2017,Motta-2018}.

The development process of IoT-based applications can be enhanced to meet new challenges by using patterns, where a pattern is a description of a set of predefined subsystems, written by experienced developers, for promoting best programming practices \cite{Fowler-2002}. In other words, when a particular type of recurring problem is solved by many developers in a similar manner and accepted by many other developers, this becomes a pattern. For example, Model-View-Controller (MVC), an architectural pattern created for web applications and implemented by several programming languages, proposes a clear separation into three layers: 1) Model: handles data processing and business rules; 2) View: renders the graphics components and notifies the Controller layer; 3) Controller: defines the application behavior regarding data input/output and application browsability \cite{Leff-2001}.The use of patterns has several advantages, such as \cite{Aksu-2018,Gamma-1994}: a) software reuse becomes easier since patterns provide vocabulary, and increase the understanding of project solutions; b) standard names become part of a widespread design language, as they eliminate the need to explain a solution to a given problem with a long description; c) patterns become a means to document software architectures. In other words, patterns contribute to software development with defined properties \cite{Gamma-1994}.

Components, connectors, and the relationship between them define a software architecture  \cite{Garlan-1993}, where connectors are often implemented with project patterns \cite{Heuzeroth-2003}. There are several IoT software solutions in the literature, but no work classifies and justifies the components in order to facilitate software architecture decisions for IoT application developers. In general, these only show an ad-hoc resolution of a specific problem, with no explanation and justification of the choices of software components and connectors. By organizing the components and connectors, as proposed in this paper, it is possible to have a clearer view of how potential components can connect and compose a software architecture to solve a specific problem.

This paper classifies architectural patterns into data ingestion, data interaction, data integration, data processing, data visualization, and data security. This classification allows the developer to have a clear view when choosing which software components are best suited for a specific IoT application, such as cities, agriculture, health, and industry. We exemplify the use of some identified components and connectors with three examples of applications in cities, buildings, and agriculture. Our primary purpose is to review the literature regarding software architectures for IoT and, therefore, propose a classification for components related to the development of IoT applications. The importance of this classification is because, as previously stated, the choice of software components and the connectors between them is an essential factor for better decision making, and also brings more clarity to viewing and understanding the software architecture. Therefore, this paper has the following contributions:

\begin{itemize}
    
\item \textbf{IoT Architectural Patterns:} A classification of existing software architectural patterns and how they relate to the inherent characteristics of IoT.
    
\item \textbf{Software Components for IoT Deployment:} Based on the classification of IoT architectural patterns, a set of essential IoT software components and connectors is presented.    
    
\item \textbf{Case Studies:} Three case studies demonstrate the contribution of patterns, components, and connectors to the literature of the area and to guide future developments of IoT-based applications. 
    
\end{itemize}

The following sections are organized as follows. Section \ref{sec:background-related-work} reviews the literature, offering a discussion of IoT-related work, and Section \ref{sec:methodology} presents the methodology used in the development of this paper. Section \ref{sec:padroes-aquiteturais-existentes} presents the classification of eminent architectural patterns related to IoT, and Section \ref{sec:componentes} presents a set of software components and connectors, based on the previous classification, which can be used to organize and improve the choices of a software developer or architect. In section \ref{sec:case-study-architecture}, there is a study on some software architectures known in the literature. Section \ref{sec:arquiteturas-referencia} shows a study with three classes of IoT applications, from which software architecture patterns are extracted from the literature. Section \ref{sec:discussion-challenges} presents a discussion of lessons learned, as well as challenges for the future, while section \ref{sec:conclusion} presents the final remarks.

\section{Background and Related Work}
\label{sec:background-related-work}
There are some widely accepted definitions of Software Architectures, such as the one given by Kruchten \cite{Kruchten-2000}, where ``software architecture encompasses the set of significant decisions about the organization of a software system.'' More formally, software architecture can be defined as the set of structures needed to reason about the software system, which comprises the software elements, the relations between them, and the properties of both elements and relations \cite{Bass-2012}. Software Architectures also play a fundamental role as a bridge between requirements and implementation.  The ISO 42010 standard \cite{ISO-42010} defines requirements on the description of system architectures, focusing on views as key integral part of the architecture description that addresses concerns of stakeholders. 

IoT introduces a new level of complexity for software development due to its inherent distributed nature with a vast amount of devices and, therefore, a new breed of software architectures is required \cite{Weyrich-2016}. The process of building software architectures for IoT involves the interaction of a variety of components with different roles. The IoT-A project proposed an architectural reference model and a preliminary set of buildings blocks to promote a fully interoperable and scalable vision of IoT \cite{Bassi-2016}. The foundation of the IoT-A Reference Model is the IoT Domain Model, which introduces the main concepts of the Internet of Things like Devices, IoT Services, and Virtual Entities (VE), as well as the relations between these concepts. The abstraction level of the IoT Domain Model also has been adopted in this work due to its concepts are independent of specific technologies and use-cases. 

Building interoperable IoT services and applications requires a set of middleware components and system development and deployment tools for rapid software development. In order to avoid developing extremely focused and vertical IoT applications not able to interact with each other, common and generic middleware services used by different application domains become necessary. Razzaque et al. \cite{Razzaque-2016} identified a variety of requirements in different categories for an IoT Middleware, which are also aligned with the functionality groups identified by IoT-A. One of these IoT requirements is to offer a means for enabling the cooperation between objects and humans and creating awareness about the surrounding environment (context awareness) in a fully connected environment. Context-aware systems can be defined as systems that adapt their behavior to the current context conditions without explicit user intervention \cite{Baldauf-2007}. Currently, with the advent of the Internet of Things and the countless applications in different areas, context-aware management has been increasingly used within the scope of big data analytics \cite{Moreno-2017,AlNuaimi-2015,Kamienski-2018-Tracker}. 

Lee \textit{et al.} \cite{Lee-Law-2017} proposed five design patterns related to the security of IoT systems, which are secure logger pattern, input validation pattern, secure directory pattern, secure adopter pattern, and exception manager pattern.  Qanbari \textit{et al.} 2016 \cite{Qanbari-2016} proposed design patterns for applications that use edge computing: 1) edge computing pattern to handle the provision of all edge devices automatically; 2) source code deployment pattern for edge devices to handle the deployment of the code to all devices connected to the IoT system; 3) edge orchestration pattern to handle the automation of creation, monitoring, and deployment of resources in the IoT Environment. Brambilla \textit{et al.} \cite{Brambilla-2017} proposed a set of design patterns for user interaction with IoT applications. Graphical user interface patterns are addressed, which involve the development of solutions for IoT, as well as an Interaction Flow Modeling Language (IFML) to express content, user interaction, and front-end behavior control of software applications. This paper recognizes the importance of design patterns for data visualization but addresses the theme more broadly.

Reinfurt \textit{et al.} \cite{Reinfurt-2016} studied many IoT solutions and extracted five lower-level design patterns for IoT devices: 1) device connection pattern on the network; 2) rule-based inference pattern, whose rule specification language exempts the user from knowing programming languages;
3) device activation trigger pattern, used to send messages to devices that are not connected to the server but can listen to the messages; 4) wake device pattern, which allows disconnecting devices from the network to reconnect them when needed; 5) remote lock and cleaning pattern, used to control stolen or missing devices. Koster \cite{Koster-2019} briefly discusses some design patterns for connected devices, IoT use cases, information models, interaction, application programming, infrastructure, and IoT security. These design patterns provide a better solution for building an architecture model for IoT. However, Koster only addressed IoT-related design patterns but did not relate design patterns to components of IoT software architectures, as presented in this paper.

Silva \textit{et al.} \cite{Silva-2013} discuss the main software architecture projects being proposed in the literature, mainly an approach to the requirements involved in each project and their respective software architectures. The authors also discuss the importance of software requirements in Smart Cities. Although this paper discusses some components, it does not classify and associate them with patterns of existing software architectures. In comparison, in this paper, our discussion is focused on components and connectors of an IoT software architecture. Yin \textit{et al.} \cite{Yin-2015} also address projects for smart cities, presenting an understanding of different smart city domains: government, citizens, business, and environment. They also analyze software architectures with particular attention to the data. They present some research challenges and propose a four-layer software architecture: data acquisition, data vitalization (the relationship between physical and virtual data), data-related services, and application domain. Yin \textit{et al.} discuss theoretical aspects of smart cities and do not contextualize practices of choosing components and connectors of a software architecture.

Ray \textit{et al.} \cite{Ray-2016} present the state of the art in IoT and the open problems associated with IoT software architectures, also discussing  essential concepts behind them. The work focuses on specific architectures of the area of IoT applications, highlights the challenges and enables future research opportunities in software architectures and IoT as a whole.  Mahmoud \textit{et al.} \cite{Mahmoud-2015} describe a three-layer architecture - perception, network, and application layers - and present different challenges related to security and IoT devices. They also present the IoT security state of the art and future work related to IoT. Here, the discussion is focused on components, and connectors of an IoT software architecture, whereas Mahmoud \textit{et al.} discuss some components, but do not classify and associate them with existing architectural patterns.

Santana \textit{et al.} \cite{Santana-2017} analyzed 23 software platforms for smart cities based on functional and non-functional requirements, and classified them into four categories: Cyber-Physical Systems, Internet of Things, Big Data, and Cloud Computing. As an outcome of this study, they proposed a reference architecture to guide the development of next generation smart cities platforms, highlighting a variety of system domains that may facilitate software development, such as urban mobility, air pollution and healthcare. While this paper has a role in the development of smart cities platforms, it follows a traditional approach based on requirements. On the other hand, here we focus on different architectural patterns for the development of IoT smart applications, not only for smart cities.

Finally, the well-known survey of Atzori \textit{et al.} \cite{Atzori-2010} presents algorithms, protocols, and solutions for IoT, as well as open challenges. They discuss different views of the IoT paradigm, according to the evolution up to 2010, describing the main classes of possible applications. However, they do not address software architectures and patterns for IoT application development because, at the time of their publication, the challenge was understanding potential applications, protocols, and technologies.

The papers mentioned above made significant contributions to literature and practice as they focus mostly on IoT devices and smart city concepts. However, they do not focus on building software architectures and development patterns for IoT applications, with a study that covers the iteration between components and their connectors. Instead, this paper addresses design patterns for developing IoT systems more broadly, including aspects such as data ingestion, interaction, storage, visualization, and processing. To the best of our knowledge, the existing literature does not cover together patterns of software architectures, components and, connectors, as well as examples of IoT applications. Our goal is to properly contextualize this area, which is likely to show significant growth in the coming years and to guide software developers to build advanced IoT applications.

\section{Methodology}
\label{sec:methodology}

Since this paper seeks to present a classification of architectural patterns required to implement an IoT solution, a methodological approach based on hands-on learning gained from real IoT usage scenarios has been adopted. In particular, this work builds on the experience gained from two international projects in tool development: a) IMPReSS \cite{Kamienski-2017-1} and b) SWAMP \cite{Kamienski-2019}.

The IMPReSS project focused on the efficient management of electricity in public buildings, but it is also possible to apply it in scenarios that aim to make society smarter. Through this project, a platform was created that allows quick development of applications for context-sensitive scenarios. This platform comprises a variety of components to render the task of developing IoT-enabled applications more straightforward, including tools for agile development of user interface, data storage with recognition, context analysis and management, mixed-criticality management, and wireless IoT communication management \cite{Kamienski-2017-1}. The experience with IMPReSS has brought two critical learnings. Firstly, arbitrary decisions of patterns, architectures, and software components do not necessarily meet the needs of IoT applications. These decisions came from experiences with developing applications with different IoT characteristics, and the resulting solutions did not deliver the expected results. For example, achieving performance compatible with IoT applications with thousands or millions of sensors required many adjustments to the architecture and connection of the components \cite{Borelli-2018}, \cite{Kamienski-2018-Tracker}. Secondly, the pattern of connection between software components has a significant influence on the performance and the scalability of solutions, which is often overlooked in the literature. For example, with the IMPReSS context-aware manager, many combinations of components and connectors have been tested to attain a performance compatible with IoT needs \cite{Borelli-2018}.

More recently, the SWAMP project seeks to develop IoT-based methods and approaches for smart water management in the precision irrigation domain and to deploy the results obtained by the project in four places: two pilots in Europe (Italy and Spain) and two pilots in Brazil. Besides, this project aims to improve precision irrigation by monitoring field status (size, growth phase) based on crop and environment, and adjusting the irrigation plan. Water management pilots aim to ensure that the technological components are flexible enough to adapt to different contexts and to be replicable to different locations and configurations \cite{Kamienski-2019}. The experience with SWAMP, still under development, is complementary to IMPReSS. Firstly, simpler architectures centered on the efficient  data distribution (for example, using FIWARE Orion Context Broker), sharing data processing and storage among different software components, add strength and scalability to solutions. Secondly, the environment where IoT applications run is inherently distributed, involving different devices and locations for processing and use. The sensors and actuators are installed on the field (farms) and communicate with intermediate elements, called IoT gateways or radio gateways, which may or may not use fog or edge computing support. Data is sent to the cloud for processing according to the application's models (irrigation) and the consequent generation of user interface services.

Hands-on experience with the development of IoT applications for smart buildings and precision irrigation has led to the need for extensive state-of-the-art study (still in an early state), experimentation with different technologies and concepts not yet fully consolidated and understanding how choices affect functionality and performance. In other words, best practices were understood and classified during the process of developing IoT applications. In general, the literature presents the use of IoT devices (sensors and actuators), device technologies (e.g., Arduino, Raspberry Pi, Gateways), wireless communication technologies (e.g., LoRaWAN), protocols (e.g., MQTT, CoAP), IoT platforms (e.g., FIWARE, AWS IoT, Google IoT) and data management and processing systems designed for applications such as online social networking (e.g., Kafka, Spark). Design choices and component connection patterns are often arbitrary, meaning that there is no apparent justification for the need, requirements, suitability, and trade-offs of existing solutions to address the challenges of the applications. The purpose of this paper is to shed some light on this discussion and provide insights to guide and justify decisions related to IoT application development projects.

From the lessons learned from these research papers, a classification for design patterns has been defined, consisting of seven classes of patterns, which should be analyzed, considered, and investigated during the structuring, design, and development of a solution for a real IoT problem.

\section{Architectural Patterns}
\label{sec:padroes-aquiteturais-existentes}
Here we explore the following aspects considered necessary for IoT software development, compiled from our experience and relevant literature in the area: 
 
 \begin{itemize}
     \item \textbf{Data Ingestion:}  specify the management of message input, which also involves output and transmission among components.  
    \item \textbf{Data Interaction:} explores how components exchange messages in a system.
     \item \textbf{Data Storage:} relates to data storage and retrieval in a system.
     \item \textbf{Data Integration:} involves computational techniques to combine data from different sources \cite{Hall-1997}.
     \item \textbf{Data Processing:} focuses on data processing, whether for decision making or transformation of data into information.
     \item \textbf{Data Visualization:} explores techniques for the visualization of information by users.
     \item \textbf{Data Security:} explores methods and techniques for enabling applications to protect their data and physical devices.

 \end{itemize}

These aspects are detailed as follows in the context of IoT software architectures.

\subsection{Data Ingestion}
\label{subsec:ingestion-data}

In most scenarios involving real applications, data input, or ingestion plays a vital role. Data ingestion is characterized by the existence of many data sources \cite{Qiao-2015}, in which processing becomes more complex as the number of data sources increases. Some of the main patterns for data ingestion are:

\begin{itemize}
\item \textbf{Multisource Extractor:} refers to the ingestion of multiple data sources efficiently \cite{Sawant-2013}. The multisource extractor pattern is recommended in scenarios where large data collections are available in different application domains, and it is necessary to investigate these data sets generally found in databases that do not follow the relational model \cite{Lengyel-2015, Huang-2014}. Generally, in large data ingestion systems, \textit{enrichers} are used to aggregate and clean the initial data. A reliable enricher transfers, validates, reduces noise, and compresses files and transforms them into a native format to deliver a representation that is easy to interpret \cite{Sawant-2013}. For example, Bashir and Gill \cite{Bashir-2016} used the concept of enrichers when storing data coming from sensors in the Hadoop Distributed File System (HDFS) with Apache Flume as the message bus.

\item \textbf{Protocol Converter:} employs a mediation component to provide an abstraction for data received from different protocol layers \cite{Sawant-2013}. The protocol converter pattern is applicable in scenarios with a wide range of unstructured data from sources using different protocols and data formats. This situation often occurs in IoT, as sensors from different manufacturers with different communication technologies transfer data in different formats. Conversion is required when data sources use several different protocols to standardize the structures of many different messages for making it is possible to analyze the information using an analysis tool. This pattern is common in IoT middleware \cite{Razzaque-2016}, which tends to convert several communication protocols transferring data from many sources, such as sensors. Marosi \textit{et al.} \cite{Marosi-2018} have created an IoT software architecture for collecting weather data, images, and soil data to enable precision agriculture. Conversion protocols allow storing information coming from these many sources.

\item \textbf{Multidestination:} used in scenarios where processing components need to transport data to many storage destinations, such as Hadoop Distributed File System (HDFS), data lakes \cite{Miloslavskaya-2016}, or real-time analytical engines \cite{Verma-2017}. The multidestination pattern is similar to the multisource ingestion pattern. Cenni \textit{et al.} \cite{Cenni-2017} created a crawler that queries Twitter API and stores the results in a storage component that feeds statistical analysis, natural language processing, and sentiment analysis.

\item \textbf{Just-in-Time Transformation:} large amounts of unstructured data are processed into batches using traditional ETL (Extract, Transform, Load) tools and methods. However, in just-in-time, data is transformed only when needed to save computing time, as described in Section \ref{subsec:etl-elt}. Colmenares \textit{et al.} \cite{Colmenares-2017} introduce a data storage system capable of ingesting sensor data at very high rates and query times suitable for large-scale smart applications. This work used global and local data structures to store data in storage components with indexes updated according to the insertions. Thus, data transformation occurs as a data set is inserted.

\item \textbf{Real-Time Streaming:} some issues require instant analysis of the data at the moment it is generated. Under these circumstances, ingestion and real-time analysis of streaming data are required. Ta-Shma \textit{et al.} \cite{Ta-Shma-2018} proposed an architecture for real-time data ingestion and historical data processing. Typically, IoT applications have a requirement to respond to real-time events based on the knowledge of past events. Real-Time Streaming design pattern is used to process events at a pace compatible with the needs of smart applications.

\end{itemize}

\subsection{Data Interaction}
\label{subsec:interaction-pattern}

Data interaction patterns describe how different components of a system interact and communicate with each other, including communication protocols. Data Interaction is different from Data Ingestion because the former focuses on the best way to establish a communication between two system components, whereas the latter aims at understanding the mechanisms used to enter data into the system. Some data interaction patterns are presented below.

\begin{itemize}

\item \textbf{Request/Response:} one of the most basic communication patterns, allowing a client to request information from a server in real-time  \cite{Tanenbaum-2006, Endrei-2004,  Alonso-2004}. Esposte \textit{et al.} \cite{Esposte-2017} proposed  InterSCity, an open-source platform  for smart cities based on microservices. Its goal is to provide a high-quality, modular, scalable, and reusable middleware infrastructure to support smart city solutions. All microservices offer communication via RESTful \cite{battle2008bridging} based on the Request/Response pattern. Almeida \textit{et al.} \cite{Almeida-2013} present the Thing Broker, a Web of Things platform that provides RESTful interfaces \cite{RESTFul} using a set of abstractions to enable communication with Twitter based on the Request/Response pattern via the Twitter Streaming API.

\item \textbf{Asynchronous Messaging:} allows software components to send messages asynchronously in real-time to other software components \cite{Tanenbaum-2006, Endrei-2004}.  Akbar \textit{et al.} \cite{Akbar-2017}  propose a generic open-source architecture based on components to combine machine learning with data processing to predict complex events for IoT applications. Apache Node-RED \cite{nodered2019} provides the ingestion of data from different sources, such as MQTT \cite{oasis2018mqtt} or RESTful API. After the data is ingested via Node-RED, it is published in Apache Kafka \cite{garg2013apache} to be accessed by machine learning algorithms . Apache Kafka is an example of an asynchronous messaging component.

\item \textbf{Publish/Subscribe:} an efficient data distribution pattern by reducing network traffic that allows a publisher to send a message only once to an intermediary, usually called broker, which in turn sends the message to the subscribers \cite{Tarkoma-2012}. MQTT is one of the most commonly used IoT protocols today to bring data from devices connected directly to the Internet through an IP address and deploys the publish/subscribe communication model. Chen and Lin \cite{Chen-2014} implemented an MQTT Proxy in their RESTful architecture, comparing latency and performance between protocols. However, they do not discuss how to implement a RESTful-like functionality with MQTT. Zyrianoff \textit{et al.} \cite{Borelli-2018} used the MQTT protocol to send data from the sensors to the data fusion module.

\item \textbf{Synchronous Messaging:} allows software components to send messages synchronously in real-time to other software components. Zyrianoff \textit{et al.} \cite{Borelli-2018} conducted a study on the impact of storing data synchronously (the application waits for a response from the database) or asynchronously (the application queues data to be stored in the database by another service and resumes its activity). In other words, whenever an application waits for a confirmation from the database that the data was stored, the message exchange between the components is synchronous.

\end{itemize}

\subsection{Data Storage}
\label{subsec:amazenamento-dados}

The need to store large volumes of data from different sources has forced databases to follow new rules of relationships and integrity, which differ from the traditional relational database management systems (DBMS). Some Data Storage patterns that can be used in IoT systems are presented below.

\begin{itemize}

\item \textbf{SQL:} Relational databases are based on query language (SQL) to define and manipulate data, which is extremely powerful: SQL is one of the most versatile and widely used options, being a safe choice and especially suited for complex queries. SQL requires the use of predefined schemas (visual and logical architectures) to determine the structure of the data before working with it. It also requires all the data to follow the same structure \cite{Beighley-2007}. 

In most situations, SQL databases are scalable vertically, allowing a performance boost when loading into a server by improving aspects such as CPU, RAM, or SSD \cite{Rautmare-2016}. Phan \textit{et al.} \cite{Phan-2014} studied different types of cloud databases, assessing and comparing databases, as well as pointing out their differences in performance, usage, and complexity. Their focus was on assessing the most common types of IoT data, with extensive experiments using four prominent databases, including MySQL.

However, relational DBMSs follow ACID rules of atomicity, consistency, isolation, and durability, which make the database reliable for its users \cite{Beighley-2007}. Storing and retrieving large volumes of data generated by IoT applications collide with ACID properties, which processes data slower than it is generated \cite{Tropashko-2007}. For addressing these new challenges brought by IoT, new data storage patterns have been used.

\item \textbf{NoSQL:} NoSQL brings together a range of databases that do not follow the relational model, and thus eliminates a fixed schema, avoids joins and facilitates scalability because, in SQL, schemas are predefined to determine the structure of the data before working with it, and all the data must follow the same structure. On the other hand, NoSQL databases have a dynamic schema for unstructured data, storing data in many different ways  \cite{Lee-2015-2}. NoSQL stands for "Not Only SQL" or "Not SQL" \cite{Han-2011}. While relational databases use SQL to store and retrieve data, NoSQL encompasses a wide range of database technologies to store structured, semi-structured, unstructured, and polymorphic data \cite{Stonebraker-2010}.

Key-value databases store data as simple key and value pairs. The keys are unique and have no restrictions. Besides, this technology does not embrace concepts such as foreign key and integrity. Key-values are suitable for parallel searches because the data sources do not have relationships between them. Due to the lack of referential integrity, integrity must be managed by the applications using the data \cite{Silberschatz-1996}.

Column-oriented databases have many columns, each with its key, for each tuple. Related columns have a column family qualifier to enable joint retrieval during a search. Since each column also has a key, these databases are suitable for fast writes \cite{Han-2011}.

Document databases store text, media, JSON, or XML data. This type of NoSQL database is suitable for cases where it is necessary to search for many documents for a specific query \cite{Han-2011}.

Graph databases store data entities and connections between them as vertices and edges, to which graph and social network algorithms and metrics may be applied, such as shorter paths and centrality \cite{Han-2011}. 
Cecchinel \textit{et al.} \cite{Cecchinel-2014} describe a software architecture that collects IoT data. In such a situation, the data is sent by the sensors and stored in storage components. This architecture faces several challenges, e.g., data storage, avoiding processing bottlenecks, sensor heterogeneity, and high productivity. To deploy the database, we opted for MongoDB, which is a NoSQL database.

\item \textbf{HDFS:} Hadoop Distributed File System (HDFS) is the primary data storage system used by Hadoop, supporting high-speed data transfer rates between compute nodes.
Initially closely linked to MapReduce, a programmatic framework for data processing \cite{Grover-2015,Bengfort-2016}, today, it is also adopted by modern big data processing systems. 

When HDFS collects data, it divides the information into separate blocks and distributes them to different nodes in a cluster, thus enabling a highly efficient parallel processing. Fault-tolerance was also a  design guideline for HDFS. The file system replicates or copies each piece of data many times and distributes the copies to individual compute nodes, placing at least one copy in a server rack different from the others. As a result, data on unavailable compute nodes can be found elsewhere within a cluster, ensuring continuity of processing while data is retrieved \cite{Grover-2015,Bengfort-2016}.

HDFS uses a master/slave architecture. In their early versions, each Hadoop cluster consisted of a single master (NameNode) that managed file system operations and supported slaves (DataNodes) that managed data storage on individual compute nodes \cite{Grover-2015,Bengfort-2016, Erraissi-2017}. Miner and Sook \cite{Miner-2012} present different design patterns involving MapReduce, including some that use HDFS to store and distribute data processing.

\item \textbf{Polyglot:} systems can use multiple types of storage, such as relational databases, files, HDFS, and NoSQL. Liu \textit{et al.} \cite{Liu-2017} proposed a model for smart cities where data management uses ingestion tools to process large amounts of data. To create this ETL-based data ingestion tool, the authors used many storage components and deployed the Polyglot pattern.

\end{itemize}

This section discusses the major design patterns for data storage. However, it is also necessary to integrate this data, whose patterns are presented in the next section.

\subsection{Data Integration}
\label{subsec:integracao-dados}

Data moves across systems but is not always in a standard format. Data integration aims to make data usable across all systems of interest so that they can be accessed and manipulated by their composing subsystems \cite{Sawant-2013}. Data integration patterns facilitate data usability by standardizing the integration process. Some data integration patterns are:

\begin{itemize}

\item \textbf{Migration:} migration means moving a dataset from one system to another. A migration contains a source system in which data resides before execution, a criterion that determines the scope of the data to be migrated, a transformation through which the data set will go through, a destination system into which the data will be inserted, and a feature to capture the migration results to know the final state versus the desired state \cite{Sawant-2013}. Migrations are essential for all data systems and are used extensively in any organization that has data operations.

\item	\textbf{Aggregation:} aggregation is the act of collecting or receiving data from multiple systems and entering it into a single system. The aggregation pattern derives its value from extracting and processing data from multiple systems into a unified application \cite{Sawant-2013}. As an outcome,  data is up-to-date at the exact moment it is needed, is not replicated, and can be processed to generate the desired dataset. The aggregation pattern is ideal for creating orchestration APIs to modernize legacy systems, especially for APIs that get data from multiple systems and then process the data into a single response. Bellini \textit{et al.} \cite{Bellini-2018} proposed a smart city architecture that includes a data aggregation layer using a multidomain ontology, creating a knowledge base for the city (some sort of expert system capable of inference).

\end{itemize}

\subsection{Data Processing}
\label{subsec-data-processing-pattern}

Data processing is a conversion of raw data into meaningful information. Data is technically manipulated to generate results that guide and help to solve a problem or improving an existing situation.

\begin{itemize}
\item \textbf{MapReduce:} MapReduce \cite{Dean-2008} is a model developed for programming parallel applications using the divide-and-conquer paradigm \cite{Cormen-2009} with two phases. In the first phase, called Map, the data is separated into key and value pairs, divided into fragments, and distributed to the compute nodes for processing. In the second phase, called Reduce, the processing performed by compute nodes by a master node \cite{Wu-2017-big-data} is combined, which generates a single response to the request made by the user. MapReduce is an adequate solution for a wide range of domains, including data mining and machine learning, social media analysis, financial analysis, image retrieval and processing, simulation, website tracking, machine translation, and bioinformatics. Currently, MapReduce is considered one of the most important parallel programming models for distributed environments \cite{Belcastro-2018}.

Apache Hadoop \cite{Hadoop} is the most widely used open-source MapReduce implementation. It can be adopted for developing distributed and parallel applications using different programming languages. Hadoop relieves developers from having to deal with classical distributed computing issues, such as load balancing, fault tolerance, data locality, and network bandwidth saving \cite{Wu-2017-big-data}.

Qureshi \textit{et al.} \cite{Qureshi-2019} propose storing data block replicas in edge components and running MapReduce to group the data processed by the edge, thus reducing the network workload of moving large data sets and reduces latency on the network.

\item \textbf{Functional:} this is a programming paradigm that treats computing as an assessment of mathematical functions and avoids changeable states or data. It emphasizes the application of functions, in contrast to imperative programming, which emphasizes changes in the program state \cite{Hughes-89, Hudak-1989}. Functional programming is becoming the emerging paradigm for distributed data (big data) processing systems such as Apache Spark \cite{Spark} and Flink \cite{Flink}, which use functional interfaces to make it easy for programmers to write data in applications in an uncomplicated and declarative manner. In functional programming, interfaces are specified as functions applied to the input of data sources. Compared to object-oriented programming, functional programming is more compact and intuitive to represent transformations based on data and applications \cite{Wu-2017-big-data}. 

Feng \textit{et al.} \cite{Feng-2018} propose an IoT-based processing system of  hydrological rainfall data using Apache Flink \cite{Flink}. Experimental results show that the processing capacity of the hydrological data processing system using Apache Flink is much higher than the traditional multilayered architecture system based on Java EE (Enterprise Edition) or pure NoSQL databases. The authors consider this solution suitable for the automation of water conservation systems.

\item \textbf{Statistical:} statistical programming refers to computing techniques that assist in data analysis. Statistical programming packages offer a wide variety of techniques for exploring large data sets and creating charts to improve understanding of the results. These packages support statistical techniques such as linear and nonlinear modeling, classification, grouping, and time series analysis  \cite{Wiley-2015}.

Nesa \textit{et al.} \cite{Nesa-2018} present an IoT architecture for detecting errors and events in a forest environment with the help of four statistical models, considering the data spatial and temporal dependencies. Simulation results show that models can effectively detect both types of outliers, with an accuracy of up to 100\% for error detection, and up to 98.51\% for event detection.

\item \textbf{SQL-Like:} NoSQL databases address several issues related to data storage and management, but in many cases, they are not suitable for performing analyzes \cite{Belcastro-2018}. Although systems based on MapReduce can handle scalability issues and decrease query times, users with little knowledge of this approach spend much time to perform simple operations such as an addition or calculating an average. To address this scenario, some systems have been developed to facilitate access to query resources of systems based on MapReduce, such as Hadoop, and development of data analysis applications using a language similar to SQL \cite{Belcastro-2018}.

Grover \textit{et al.} \cite{Grover-2015-sql-like} benchmarked several big data SQL-type technologies on the Hadoop Distributed File System (HDFS) used in clinical trial databases.

\item \textbf{Data Fusion:} this is a technique that combines data from multiple sources and associates them to increase accuracy and improve inferences when compared to the information obtained from only one data source \cite{Hall-1997, Klein-1999, Llinas-1989}. Zyrianoff \textit{et al.} \cite{Borelli-2018} and Kamienski \textit{et al.} \cite{Kamienski-2015-context} used data fusion to combine data from temperature sensors. This fusion was based on the average in a smart city scenario. By merging the data, it was possible to preprocess data before sending it to the next component of the software architecture in the smart city scenario.

\item \textbf{Rule Engine:} this is a solution for managing business rules in constant evolution by separating the knowledge of the business rules from their deployment in the system of interest \cite{Nagl-2006}. Zyrianoff \textit{et al.} \cite{Borelli-2018} and Kamienski \textit{et al.} \cite{Kamienski-2015-context} used a business rule inference engine to verify what action an actuator inserted in a smart city scenario should take based on the data processed by the data fusion component. Thus, business rules can be changed at any time because they have been separated from the logical deployment of the system to manage a smart city scenario.

\end{itemize}

Bashir \textit{et al.} \cite{Bashir-2016} analyzed large amounts of smart building data using sensors to measure the concentration of oxygen and other gases in indoor environments. The authors propose a three-layer framework: IoT sensors, data management, and data analysis, and used a processing component analyzes data stored in HDFS in real-time. If the oxygen concentration level captured by the sensor is within a threshold preset as comfortable, no action is required. Otherwise, an oxygen pump is activated until oxygen levels are within a comfortable limit for the user.

\subsection{Data Visualization}
\label{subsec:visualizacao-dados-pattern}

Data visualization has changed in recent years, evolving from a simple visual representation into analysis techniques to aid the interpretation of data in a broader sense. 

\begin{itemize}
\item	\textbf{Mashup View:} Mashup View is used to maximize query performance by storing a view of mashups aggregated in the storage layer \cite{Abiteboul-2008}. This data visualization pattern reduces analysis time by aggregating the result into a storage layer.  Blackstock and Lea \cite{Blackstock-2012} propose a platform that provides a simple way for users to find, control, view, and share device data. This platform is based on mashups. Soukaras \textit{et al.} \cite{Soukaras-2015} present a development platform with a toolkit called the IoTSuite, consisting of an editor, a compiler, and a development module connected to a mapper and a link connector. This development platform is based on mashups to connect all IoTSuite components.

\item	\textbf{Portal:} an organization that has a corporate portal can follow this pattern and reuse it for big data visualization.  Merlino \textit{et al.} \cite{Merlino-2014} propose an extension of OpenStack in a smart city scenario to manage sensors, analyze data, and provide a real-time sensor and data visualization panel.

\end{itemize}

\subsection{Data Security}
\label{subsec-data-security-pattern}

Implementing security measures is critical to ensure the proper operation of networks carrying data from IoT devices. Several challenges prevent the protection of IoT devices and end-to-end security in an IoT environment. Security was not always considered a top priority during the product design phase. Besides, since IoT is a new and expanding market, many product designers and manufacturers are more interested in getting their products to the market quickly rather than taking the necessary steps to ensure security since the design phase \cite{Conti-2018, Andrea-2015}.

A significant vulnerability in IoT systems is the use of standard or embedded passwords, which may lead to security breaches. Even if passwords are changed, they are usually not strong enough to prevent hacking. Another common problem faced by IoT devices is the restriction of computational resources to implement strong security, which generates weaknesses across multiple devices. For example, sensors that monitor soil moisture or air temperature cannot handle advanced encryption or other security measures. Also, because many IoT devices are configured and forgotten - that is, they are put in the field and left there until the end of their life - they rarely receive security updates or patches \cite{Conti-2018,Zhou-2017-security, Choudhury-2017}.

The lack of industry-accepted standards also undermines security in IoT. Although there are many proposed IoT security approaches, there is no consensus on a preferred one. Large companies and organizations may have their specific standards, while specific segments, such as industrial IoT, have proprietary and incompatible industry-lead standards. The diversity of standards makes it difficult not only to protect systems, but also to ensure interoperability between them \cite{Conti-2018}. Some security patterns proposed by the literature are presented as follows:

\begin{itemize}

\item \textbf{Authentication:} authentication enables the integration of different IoT devices and their deployment in many scenarios such as smart cities and smart agriculture. Authentication involves validation between peers of IoT devices before exchanging information to ensure that the data source is legitimate, i.e., devices of interest to the application \cite{Hafiz-2007, Alaba-2017}. Gubbi \textit{et al.} \cite{Gubbi-2013} focused on a standard authentication scheme for IoT between different layers and terminal nodes. The scheme is based on hashing and extraction of shared elements to prevent interference attacks. This scheme essentially provides an adequate security solution for authentication in IoT. The extraction procedure comprises some irreversible properties that guarantee security in the IoT domain.

\item \textbf{Authorization:} authorization involves allowing access rights to resources such as sensors or data. Data should be safe and accessible only to authorized users and systems. Gaur \textit{et al.} \cite{Gaur-2015} proposed the authentication of IoT sensor nodes that relies on a unique coding request and response scheme. The scheme uses a preshared matrix, applying a cipher of a variable when the communication involves many parties. Each communication (message exchange) between the parties is encrypted using a node key and identifier with a timestamp.

\item \textbf{Physical Security:} there is a concern about how to protect memory software and its vulnerabilities at runtime. Solutions to this problem may have to consider the specific programming languages used by IoT devices, such as the use of TinyOS \cite{Nehme-2010}. Using software management such as patching, software firmware, and remote updates can help to physically protect IoT devices \cite{ Hafiz-2007, Alaba-2017}.

\end{itemize}

Data privacy is another critical issue for IoT business procedures, and practical solutions remains a challenge \cite{Botta-2016}. The privacy of user data must be ensured by design as users need maximum protection for their personal information. Transferring and ensuring data privacy between different nodes on a heterogeneous IoT is a challenging problem because different network nodes have different trust criteria \cite{Eschenauer-2002}. The European General Data Protection Regulation (GDPR) adds urgency to the need for providing strict privacy guarantees to users of IoT applications \cite{Wachter-2018}.  Lu \textit{et al.} \cite{ Lu-2017} propose aggregation of data for computing to perform source authentication on devices at the network edge in order to pre-filter false injected data.

The previous topics described subclasses of IoT software architectural patterns involving IoT. Based on this classification, the next section presents a set of components and connectors that can help software developers and architects to design a solution to a problem involving IoT. This classification allows developers to choose more clearly the components they will use. In the literature, there is a set of articles that propose the solution to a problem and mention which patterns they used. However, most authors only describe the architecture and components chosen, not mentioning, or barely mentioning, the grounds for such a choice. Thus, although new IoT application developers find references to architectures, patterns, and components used in different projects, no compilation nor classification simplifies learning about the development of applications for IoT. The idea here is to help developers choose components for these architectures.

\section{Components for IoT Software Architectures}
\label{sec:componentes}

In this section, a set of components and connectors illustrates the design of IoT software architectures according to the pattern classification introduced in Section \ref{sec:padroes-aquiteturais-existentes}. Here, a component is understood as an independent element, which can be replaced, but is significant because it has a clear function in its specific context. We chose an IoT problem from a scenario based on Kamienski \textit{et al.} \cite{Kamienski-2015-context}, in which a public building has sensors and actuators that send messages to a context manager, where it can fuse data and infer rules that dictate how the actuators behave, to illustrate some relevant components.

\subsection{Devices} 

In general, devices are sensors, actuators, and gateways, including mechanical, electronic, and computational components. Electronic devices that house sensors and actuators are known as constrained nodes and are classified into two categories: microcontroller-class devices and general-purpose devices \cite{Bormann-2019}. Microcontrollers often include RAM and on-chip code storage, offer limited support for general-purpose operating systems, and are generally used to deploy sensors and actuators. General-purpose devices, such as Raspberry Pi, often have RAM and SSD storage on separate chips, offer support for general purpose operating systems, and are used to deploy IoT Radio Gateways \cite{Farrell-2018}.

In IoT scenarios, sensors are data sources of ingestion patterns presented in section \ref{subsec:ingestion-data}, particularly of the Multisource Extractor Pattern. They are essential components for any IoT scenario, as this component captures data to be processed and transformed into information. Actuators receive a command and execute an action in the context of the applications. For example, a lamp receives a command to turn on the light.  The use of sensors, actuators, and gateways, as well as the constrained nodes that implement them, is a pattern applicable to all IoT applications.
 
In addition to microcontrollers and general-purpose devices, IoT applications use a wide variety of equipment that cannot be strictly classified as restricted nodes, such as smartphones, laptops, desktops, and servers \cite{Bormann-2019}. These devices also have restrictions, but at different levels than constrained nodes and are used for different functions in an end-to-end view of a system.

\subsection{Data Fusion}
\label{subsec:fusao-dados}

The definition of data fusion has been the subject of intense discussion among authors \cite{Nakamura-2007}, but, in general, it is related to the process of integrating multiple data sources to generate information that is more consistent, accurate and useful than it would be if generated by only one single source. Thus, the information is of better quality or more relevant \cite{Castanedo-2013,Klein-1993}, where quality and relevance depend on the application. Examples of software components that can be used to perform data fusion are Esper \cite{Esper}, Apache Flink \cite{Flink}, and Apache Spark \cite{Spark}. Esper was used in Kamienski \textit{et al.} \cite{Kamienski-2015-context} to process data from sensors and report the average temperature of a classroom so that a Reasoning Engine could process actions on the devices, which, in this case, were air conditioners.

Some components, such as Apache NiFi, perform data fusion in addition to their primary function, which, in this case, is data pipeline. This component can be associated with the Data Processing pattern, described in section \ref{subsec-data-processing-pattern} as it processes data and transforms it into information, and may reveal patterns in the analyzed data.

\subsection{Reasoning Engine}
\label{subsec:motor-inferencia}

A rule inference engine is a component that executes one or more business rules in a running production environment. The business rules system allows company policies and other operational decisions to be defined, executed, and separated from the application code. Typically, inference engines support rules, facts, punctuation, mutual exclusion, preconditions, and other functions. The most widely used inference engines are JBoss Drools \cite{Drools}, OpenRules \cite{OpenRules}, and ThingsBoard \cite{ThingsBoard}. This component can be associated with the Data Processing design pattern presented in section \ref{subsec-data-processing-pattern}  and used as an example in works Kamienski \textit{et al.} \cite{Kamienski-2015-context} and Pramudianto \textit{et al.} \cite{Pramudianto-2016}. It can be physically located in the cloud on machines with high processing power or edge devices with low processing power. For example, this component was used by authors Kamienski \textit{et al.} \cite{Kamienski-2015-context} to process information coming from the fusion component and to take some action, such as turning off a device.

\subsection{Throttle}
\label{subsec:throttle}

In the context of web services, the throttle is a component that limits connections arriving at the web server \cite{Henderson-2006}. In the context of IoT, the throttle is used to prevent large streams and data from being sent to the next components. When a component sends multiple identical messages to another component, a component with the throttle function may prevent the next component from becoming overloaded. It can be used to implement design patterns related to data analysis presented in section \ref{subsec-data-processing-pattern}.

This component can be associated with the Data Processing design pattern presented in section \ref{subsec-data-processing-pattern}. Hiromoto \textit{et al.} \cite{Hiromoto-2017} discussed security in IoT as chain risk management. A throttle-based software architecture manages incoming and outgoing messages according to the security alerts generated by the security component. Zyrianoff \textit{et al.} \cite{Borelli-2018} saw the need for a throttle component in IoT scenarios because vast amounts of messages are sent from the fusion component to the reasoning engine component, where it processes and sends several actions to devices, which are often repeated and overload these devices. Thus, the throttle is a component that limits the messages sent from the inference engine to the devices. Although this component is not implemented by Zyrianoff \textit{et al.} \cite{Borelli-2018}, the authors are aware of the utmost importance of using this component through hands-on experiences in the IMPReSS Project.

\subsection{ Ingestion Components}
\label{subsec:ingestao-dados-componente}

This component is responsible for the process of obtaining, importing, and processing for later use in a data repository. Data, especially unstructured data, is moved from the location it originated to another component. This process usually involves altering individual files by editing their content and formatting their structure. This component can be associated with the Data Ingestion architecture pattern presented in section \ref{subsec:ingestion-data}. 

Data can be transferred in real-time or ingested in batches. When data is ingested in real-time, the data reaches its destination almost immediately after leaving the source. When data is ingested in batch, data blocks are consumed within a time interval. 

As IoT devices increase, volume and variation of data sources are expanding quickly. Thus, extracting data for use by a target system is a significant challenge in terms of time and resources. Other issues faced by data ingestion components are: 1) several data sources are being born with data formats different from the relational model and these formats are changing at a great speed and often without warning; 2) in the face of different data formats, ingesting them at a reasonable speed and processing them more efficiently to make better business decisions; 3) detection and capture of changed data due to the semi-structured or unstructured nature of the data and to the low latency required in some real IoT scenarios \cite{Kaisler-2013, Marjani-2017}.

This data ingestion component was used in the works of  Kamienski \textit{et al.} \cite{Kamienski-2015-context} together with an MQTT Broker to ingest the messages coming from the sensors and to send these messages to the data fusion component.

\subsection{ Interaction Components}
\label{subsec:transmissao-dados-comp}

This component is responsible for taking data from one component to another, which implements the interaction patterns described in subsection \ref{subsec:interaction-pattern}.

In systems with multiple components, communication between them is usually through messages, which are transferred following some connector patterns. Thus, the connector is an essential component of software architectures and belongs to interaction patterns. Here are some types of connectors that can be used. All connectors presented here can implement the Publish/Subscribe, Asynchronous Messaging, Synchronous Messaging, and Request/Response interaction patterns presented in section \ref{subsec:interaction-pattern}.

\subsubsection{Type  of Connectors}
\label{subsec-connectors-type}

Components are critical elements of a software system, but how they are connected can significantly affect the performance of a set of components in a scenario. The most common manner of communication between components is through message exchange. For two components to communicate, they must use a communication component. For example, a sensor to communicate with the data fusion component needs a communication component. Many components communicate through APIs, for example, a data fusion component communicating with a rule inference component. Some well-known manners of connecting components are:

\begin{itemize}
    \item \textbf{Serial Connector:} sends only one message from one component to another at a given time (Figure \ref{fig:serial-conector}) and is therefore suitable for small volumes of messages

\begin{figure}[ht]
 \centering

 \includegraphics[scale=2.5]{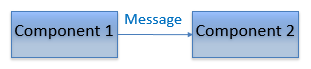}
 \caption {Serial connector: allows the transmission of only one message between components at a time. }
 \label{fig:serial-conector}
\end{figure}

 \item \textbf{Parallel Connector:} sends multiple messages between components in the same period (Figure \ref{fig:parallel-conector}), usually implemented across multiple threads, which can be created on demand. It is also possible to create initially a thread pool that are already active to perform data transfer. When threads are created on demand, some messages may experience slight communication delays between two components because the operating systems and connector applications are creating this thread. The creation of thread pools at startup time generates a slight delay in the application initialization, which is compensated by the faster use of threads afterward.

\begin{figure}[ht]
 \centering
 \includegraphics[scale=2.5]{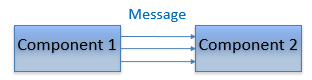}
 \caption{Parallel Connector: allows one component to send multiple messages to another connector at the same time}
 \label{fig:parallel-conector}
\end{figure}

\item \textbf{Producer-Consumer:} uses the computing technique known as Producer-Consumer to send messages between components (Figure 3). Unlike serial and parallel connectors, this connector makes the relationship asynchronous, so that variations in the arrival rate of the data at the producer are not blocked by a slower consumer service time. A queue buffers messages between the Producer and the Consumer. The Producer  stores messages in the buffer, and the Consumer retrieves them when it is ready.

\begin{figure}[ht]
 \centering
 \includegraphics[scale=2.2]{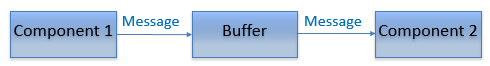}
 \caption{Producer-Consumer Connector: asynchronous transmission of messages from one component to another.}
 \label{fig:producer-consumer}
\end{figure}

\item \textbf{Data Pipeline:} uses a high-performance communication component to take messages from one component to another (Figure \ref{fig:data-logistic}), recommended in scenarios that require fault tolerance, or need to take messages to other components with delivery guarantees. The most well-known Data Pipeline connectors are Apache Kafka \cite{Kafka}, Apache NiFi \cite{Nifi}, Apache Flume \cite{Flume,Shreedharan-2014}, and Apache Flink \cite{Flink}. They make it easy to transport data between multiple software components, adding features such as fault tolerance and guarantee that the recipient will receive messages.

\begin{figure}[ht]
 \centering
 \includegraphics[scale=1.3]{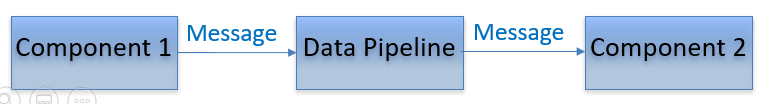}
 \caption{Data Pipeline Connector: uses a high-performance communication component to take messages from one component to another.}
 \label{fig:data-logistic}
\end{figure}

\item \textbf{API:} Application Programming Interface (API) is a set of routines and standards set and documented by an application whose functionalities can be used by other applications without the need to know implementation details, for allowing interoperability between applications. RESTFul is currently a widely used standard for implementing web services APIs, based on the REST (Representational State Transfer) software architecture style \cite{RESTFul}. Other standards exist, such as GraphQL \cite{GraphQL}, which implements APIs with a query language.

\end{itemize}

Kamienski \textit{et al.} \cite{Kamienski-2015-context} used a data interaction (connector) component to transfer messages from one data fusion component to a Reasoning Engine component. To implement this communication, we used the producer-consumer connector detailed in section V-F1. This connector creates a thread pool when the application is created, in which it receives the messages and buffer them for another thread pool to consume.

FIWARE IoT Agent is a component that transforms data coming from sensors (and going to actuators) into the NGSI standard using different protocols, such as Ultralight 2.0 \cite{Ultralight} or LoRaWAN \cite{LoRaWAN}. Ultralight 2.0 is a lightweight text-based protocol for restricted devices and communications where device bandwidth and memory may be limited \cite{Fiware-IoT-Agent}.

\subsection{Storage Components}

This component is responsible for storing the information of a system in many ways, such as SQL or NoSQL files, and databases. Regarding the design patterns presented in subsection \ref{subsec:amazenamento-dados}, IoT systems mainly use NoSQL databases due to high volume, variety of formats and speed of data insertion. Several works, such as Cai \textit{et al.} \cite{Cai-2017}, Cai \textit{et al.} \cite{Cai-2014}, Cure \textit{et al.} \cite{Cure-2012}, Zhu \textit{et al.} \cite{Zhu-2013-storage}, Ma \textit{et al.} \cite{Ma-2012}, Mallapuram \textit{et al.} \cite{ Mallapuram-2017} used storage components in their solutions. 

These solutions involve storing data schemas for IoT semantics, logs, and dataset storage, as well as storing only information from services and entities of interest. Jiang \textit{et al.} \cite{Jiang-2014} used NoSQL storage to save logs and data from RFID devices. Orion is a core component of the FIWARE platform \cite{Fiware} that acts as a data distributor, managing data context life cycle, and can also be used as a temporary storage module for IoT entities \cite{ Fiware-Orion}.

\subsection{Data Analytics}

This component is responsible for using artificial intelligence, data mining, and machine learning techniques to analyze large amounts of data and provide information, which implements analysis patterns presented in section \ref{subsec-data-processing-pattern}. There are several components available that perform both open source and commercial data analysis functions. An example is FIWARE Cosmos, used to analyze data in batches or stream to gain insight into that data by revealing new information that was hidden  \cite{Fiware-Cosmos}. Cosmos is a wrapper that interconnects FIWARE components with existing frameworks for big data analytics, especially Apache Flink. In IoT systems, time series databases are critical to store and process data, subsequently due to the constant arrival of data from large numbers of sensors \cite{Quantum-Leap}. The FIWARE platform offers the Quantum Leap component with time series database storage (currently CrateDB \cite{CrateDB}) and provides an NGSI-based API for data entry and query.

\subsection{Visualization Components}

Based on the pattern presented in section \ref{subsec:visualizacao-dados-pattern}, visualization components are responsible for displaying data in a user-friendly manner and particularly crucial for system managers and administrators.

Using visualization components allows us to display assessment metrics in a clear and organized manner automatically and allows the user to switch between different contexts that require different subsets of sensors (e.g., comfort and energy saving) installed in the IoT scenario.

Ji \textit{et al.} present a systemic analysis of requirements related to component visualization from the smart city perspective. Besides, they propose a new visual IoT architecture known as A-VIoT. The proposed system includes six main components: a) smart insight to detect complex environments; b) smart video analytics to reduce the amount of visual data; c) software-defined video to generate elastic visual streams; d) flexible controls to produce an ideal adaptation; e) cost-effective transmission to improve resource utilization, and; f) crowd coordination to improve cooperation performance \cite{Ji-2019}.

\section{Generic Architectural Patterns} 
\label{sec:case-study-architecture}

This section presents some examples of generic software architectures that fit the design patterns described here and can be used for the development of smart IoT applications.  

\subsection{ETL vs. ELT}
\label{subsec:etl-elt}

Extraction, Transformation, and Load (ETL) technologies and processes have emerged with the data warehouse concept and have now reached great maturity, remaining as the appropriate technique for Business Intelligence (BI) and Analytics solutions \cite{Kimball-2004}. The extraction phase is characterized by retrieving raw data from a set of unstructured data and migrating it to a temporary repository. The transformation phase structures, enriches, and converts raw data into a different content. Finally, the loading phase is the ingesting of structured data into a repository where it will be processed by analysis tools \cite{Bansal-2015,Vassiliadis-2009}. Figure 5a) shows the data stream for ETL.

\begin{figure}[ht]
 \centering

\includegraphics[scale=2.0]{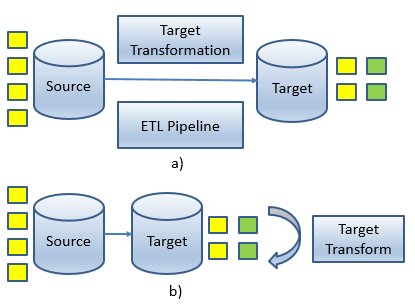}
 \caption{a) ETL is a software architecture with three phases: 1) Data extraction; 2) Data transformation according to business rules; 3) Data loading or ingestion for storage elements. b) ELT inverts ETL Transform and Load phases.
 }
 \label{fig:etl-elt}
\end{figure}

Although the ETL process offers a suitable solution for different applications, it also generates some problems of its own. An ETL process allows the Extract and Load phases to be performed at different times according to the source and destination maintenance windows so that neither source nor destination will be idle at all. With the emergence and widespread use of NoSQL databases and cloud technologies that ensure elasticity, availability, and high throughput, data can be loaded into data lakes, rendering it available to different data consumers and applications \cite{Fang-2015}.

Thus, the ELT approach (inverting the Transform and Load phases) provides an alternative to ETL. Instead of transforming data before loading it, ELT leverages the target system to do the transformation. The data is copied to the destination and then transformed there. Figure \ref{fig:etl-elt}b) shows this paradigm inversion in more detail.

ELT does not have a transformation mechanism because this work is carried out by the target system. On the other hand, in cases where the target system is not powerful enough for ELT, ETL may be more advantageous. In IoT scenarios, data from some entities can often be stored in a data lake or legacy system. A data lake is a data repository that keeps the data in its raw form without the need to worry about the structure of the data being ingested and stored \cite{Miloslavskaya-2016}. The ELT tool can help extract data from these legacy systems to be processed by analytical processing components.

Both ETL and ELT are architectures for data ingestion, as described in Section \ref{subsec:ingestion-data}. In the extraction phase, both software architectures use components of the data ingestion pattern, especially the multisource extractor pattern. In the transformation phase, both use components of the Data Processing pattern to adjust the data according to the need of the application that will be used for ETL. In the load phase, a component of the Data Storage pattern is used to store data and components of the Data Interaction pattern, to move data to the Load phase.

The sheer volume of unstructured data generated by online social network services and their requirement for real-time updating has led to the need for new scalable data management architectures. Two examples of architecture stand out: Lambda and Kappa.

\subsection{Lambda Architecture}
\label{subsec:arch-lambda}

Lambda is an architecture for processing large amounts of data that unifies online and batch processing into a single structure to balance latency, stream, and fault tolerance \cite{Marz-2013}. This pattern is suitable for applications with delays in data collection that need to show in in dashboards afterward \cite{Marz-2013,Kiran-2015}. Lambda architecture also allows processing datasets in batch, aiming at finding behavioral patterns according to the application needs.

\begin{figure}[ht]
 \centering
 \includegraphics[scale=2.0]{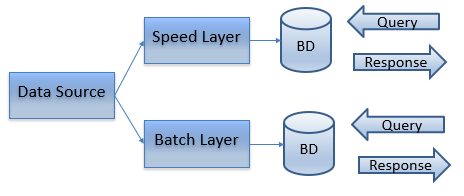}
 \caption{ Lambda architecture. This architecture has three layers: 1) Batch processing layer. 2) Real-time processing layer. 3) Visualization layer. Adapted from  Kiran \textit{et al.} \cite{Kiran-2015}.
  }
 \label{fig:lambda-arch}
\end{figure}

Figure \ref{fig:lambda-arch} shows the essential components of the Lambda architecture, with three layers: 1) Batch layer to pre-compute large amounts of data; 2) Speed layer to minimize response latency, performing calculations as data arrives; and 3) Service layer to view query results over data processed \cite{Kiran-2015}. Lambda uses data ingestion patterns, especially the multisource pattern, to power Speed and Batch layers, data storage patterns to store results of data processing, and communication and visualization patterns to access the data in the Service layer.

This architecture allows developers to optimize their data processing costs by understanding which parts of the data need to be processed in real time or in batches \cite{Marz-2013}. However, the need to develop and maintain two different codes for Batch and Speed layers requires more work from the development team, increasing the complexity of the solution [45]. Nevertheless, Lambda is suitable for big data problems, especially when processing data from sensors or another source that sends data continuously. In such cases, the Speed layer can detect anomalies in the data, and this verified data can then be stored in databases. Finally, data can be periodically processed in batch  (e.g., once a day, week, or month) to study and extract behavioral patterns.

\subsection{Kappa Architecture}
\label{subsec:arch-kappa}

Kappa architecture focuses only on data processing as a continuous stream, unifying codes of Batch and Speed layers of the Lambda architecture \cite{Kreps-2014}. Although proposed as an alternative to the Lambda architecture that solves the problem of duplicate code, Kappa has specific use cases and does not replace Lambda in all scenarios. In Kappa, incoming data is processed by a Streaming layer, and the results are placed in the Service layer for queries. The idea of Kappa architecture is to handle real time data processing and continuous reprocessing in a single stream processing engine. Reprocessing occurs from the stream. If the source code changes, a second stream process repeats all previous data through the latest real-time engine and will overwrite data stored in the presentation layer \cite{Lin-2017}.

\begin{figure}[ht]
 \centering
 \includegraphics[scale=1.8]{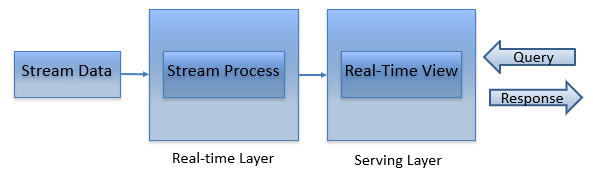}
 \caption{Arquitetura Kappa, composta por apenas uma camada de processamento}
 \label{fig:kappa-arch}
\end{figure}

This architecture attempts to simplify by maintaining just one source code base rather than managing one for each data batch and by accelerating the layers in Lambda Architecture (section \ref{subsec:arch-lambda}). The disadvantages of Kappa are related to the need to process data in a stream that is not suitable for all cases, such as handling duplicate events, cross-referencing events, or maintaining order operations that are generally easier to carry out in batch processing.

Kappa architecture uses data ingestion patterns to feed the Streaming layer, data storage patterns to store results of the data processing, and communication and visualization patterns to access data in the Service layer.

Lambda and Kappa are software architectures related to data processing, as discussed in section \ref{subsec-data-processing-pattern}. Despite the difference in how these architectures do the processing, in a nutshell, they process data and deliver this processing through visualization patterns described in section \ref{subsec:visualizacao-dados-pattern}.

\subsection{Data Analytics}
\label{subsec:analise-dados}

Data analysis from a big data perspective is different from a traditional analysis because it involves many types of unstructured data and generally related to text analysis and natural language processing \cite{Sawant-2013}. Data analysis is part of a design pattern known as pattern recognition using machine learning algorithms. Pattern recognition can be defined as the classification of data based on knowledge already gained or statistical information extracted from patterns and their representation \cite{Bishop-2006}. Statistical methods of pattern recognition have been widely applied in the field of artificial intelligence. Successful applications of these methods in the field of computer vision include extracting low-level visual information from visual images, edge detection, extraction of information of shade shapes, object segmentation, and object labeling \cite{Golden-2001, Jain-2000}.

Among the numerous applications of pattern recognition, data analysis has increasingly sophisticated methods for discovering complex structural regularities in large data sets, used in many fields such as social and behavioral sciences. Classical statistical pattern recognition techniques are used, such as factor analysis, principal component analysis, cluster analysis, and multidimensional scaling techniques. More sophisticated methods for statistical pattern recognition, such as artificial neural networks and graphical statistical models, form the basis of relevant tools for detecting structural regularities in data collected by social and behavioral scientists. Among pattern recognition techniques, machine learning provides systems with the ability to learn and improve from data without being explicitly programmed automatically. Machine learning focuses on developing computer programs that can access data and use it to learn for themselves \cite{Golden-2001, Jain-2000}.

The process of applying data analysis methods to specific areas involves defining data types (such as volume, variety, and velocity), data models (such as neural networks, classification, and clustering methods), and using efficient algorithms that match the characteristics of the data \cite{Mahdavinejad-2018}. What makes IoT data processing a challenge is: 1) data characteristics: IoT generates a massive volume of data at a very high speed, with varying formats. Data frequently is raw, with low level of abstraction, which makes it difficult to analyze it. Semantic techniques may be used to improve IoT data analyzes \cite{Roffia-2018}, which may require more effort to deal with data volume, velocity and variability; 2) data privacy: protecting  privacy is crucial because data collection processes may include personal, business and other sensible data; 3) algorithms: finding the best  model that fits the data is one of the most important issues for pattern recognition and better analysis of IoT data; the results yield by these models and algorithms may be affected by noise, as well as it may be difficult to interpret the results\cite{Mahdavinejad-2018}.

Cheng \textit{et al.} \cite{Chen-2015} introduced a systematic method for reviewing data mining knowledge and techniques in the most common IoT applications. In this study, they reviewed some data mining functions, such as classification, clustering, association analysis, and time series analysis in IoT scenarios. The authors also assigned more data mining methods to each type of IoT application and suggested a new data mining application using open source software.

Tsai \textit{et al.} \cite{Tsai-2014} conducted research to address some of the challenges in preparing and processing data for IoT using data mining techniques. The authors explain about IoT data and the challenges in this area, such as building mining models and algorithms. These are some of these challenges: 1) to show that the data chosen to be processed will solve the IoT problem in question; 2) to choose the best data analytics algorithm according to the data characteristics \cite{Tsai-2014}.

Li \textit{et al.} \cite{Li-2018} present how to apply deep learning techniques to the IoT environment to improve learning performance and to reduce network traffic. The authors prepared an elastic model compatible with different learning models. Experimental results show that the proposed solution outperforms other IoT optimization methods.

\section{Architectural Patterns for IoT Smart Applications}
\label{sec:arquiteturas-referencia}

This section use categories, components, connectors, and architectures introduced before to provide three examples of smart applications for cities, buildings, and agriculture.

\subsection{Parking Management for Smart Cities}

\begin{figure}[ht]
 \centering

 \includegraphics[scale=2.3]{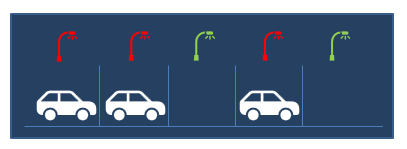}
 \caption{Smart parking. A big challenge of this scenario is to manage cars to park in the parking spaces automatically.  }
 \label{fig:smart-parking}
\end{figure}

In today's cities, finding an available parking space is always tricky for drivers, and tends to become even more difficult as the number of cars on the streets increases. This situation is an opportunity for smart cities to take action to increase the efficiency of their parking resources, leading to a reduction in parking times, traffic jams, and accidents. Problems related to parking and traffic jams could be solved if drivers could be informed in advance about the availability of parking spaces at and around the intended destination. Figure \ref{fig:smart-parking} shows an example of a smart parking scenario.

\begin{figure}[ht]
 \centering
 \includegraphics[scale=1.8]{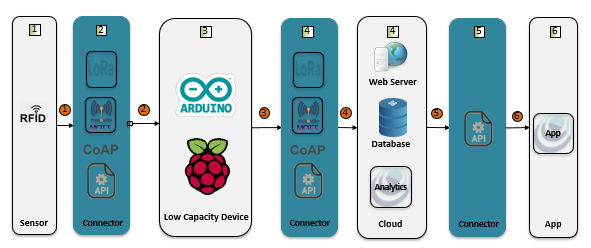}
 \caption{Reference architecture for smart parking applications. A presence sensor sends data on car occupancy to a low capacity computing component using an MQTT-based connector. The low capacity device processes data, such as the number of spaces in parking regions, and sends the information to a large-capacity processing server that typically is in the cloud using a RESTful API. In the cloud are also web and database servers. Data related to the distribution of spaces (occupied or available), and payment for usage, is processed in the cloud. Data analysis techniques are used to find patterns or suggest spaces parking closer to the user's destination.}
 \label{fig:arch-ref-parking}
\end{figure}

This type of application is a traditional problem in large cities.  Khanna \textit{et al.} \cite{Khanna-2016} and Pham \textit{et al.} \cite{Pham-2015} dealt with this issue with a software architecture pattern, as presented in Figure \ref{fig:arch-ref-parking}. A device, typically a presence sensor, checks if a car is in a parking space sends messages to a communication component, usually an MQTT broker. Components with low computing capacity, such as Raspberry Pi, subscribe to the MQTT broker and consume data from the sensors, calculate the number of parking spaces, and display the results on a dashboard. Data processed in the low capacity computing component is sent to a cloud-based high-processing server that calculates the payment for the use of the parking space and the distributions and occupancy mode between cars and parking spaces. Finally, this data is returned to the low capacity computing component.

This IoT application uses the following patterns:

\begin{itemize}
    
    \item 	Data Ingestion: Real-Time Streaming Pattern to process and distribute which users use which parking spaces
    \item 	Data Interaction: Publish/Subscribe Pattern to establish communication between sensors and low capacity processing components
    \item 	Data Storage: SQL to store data coming from sensors and their respective processing and the NoSQL pattern to manage the geographic positioning of the parking spaces
    \item 	Data Visualization: Pattern Portal to design an information portal for system administrators and managers.
    \item 	Connectors: Light Interaction to connect parking space sensors to the MQTT broker; API to connect the cloud to management applications

\end{itemize}

\subsection{Energy Efficiency Management for Smart Buildings}

\begin{figure}[ht]
 \centering
 \includegraphics[scale=2.0]{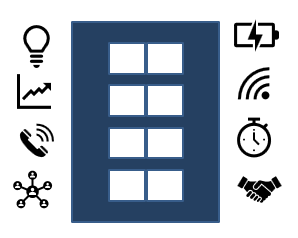}
 \caption{Building using smart energy management. The challenge is to manage a building semi-automatically, in such a way that users do not perceive the management actions.}
 \label{fig:smart-building}
\end{figure}

\begin{figure}[ht]
 \centering
 \includegraphics[scale=2.2]{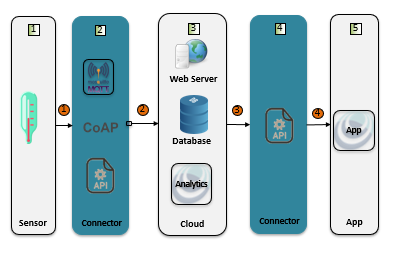}
 \caption{Reference architecture for power management applications for smart buildings. A presence sensor sends data of room temperature and presence of people inside the building to a component of high processing capacity, located in the cloud, via an MQTT based connector. The cloud hosts web servers, databases, data fusion components, and inference engines for decision making. Data analysis techniques find user behavioral patterns and change energy system behavior in a predictive and more efficient manner.}
 \label{fig:arch-ref-energy}
\end{figure}

The importance of energy in the contemporary world is growing steadily, but there are still many sources of inefficiencies in its management, such as public buildings. The path to transforming public spaces into smart environments faces several challenges, such as energy management in buildings, which are designed to automate lighting and HVAC (heating, ventilation, and air conditioning) applications \cite{Kamienski-2017-1}. Figure \ref{fig:smart-building} shows an example of a smart building scenario. Some authors have addressed the problem of power management in smart buildings \cite{Kalluri-2018, Borelli-2018, Kamienski-2015-context, Kamienski-2017-1, Minoli-2017}. Many proposed solutions for building energy management require an element of high computational power to process, store, and infer data. Figure \ref{fig:arch-ref-energy} shows a summary of this reference architecture: sensors send messages to a cloud server with high computing capacity, which in turn is responsible for storing, processing, and inferring decisions about changes in context, which may be a turn on/off and increase/decrease some equipment or device.

Existing energy management solutions for smart buildings use context-sensitive techniques. The architecture of these systems consists of sensors that send messages to a server that preprocesses and fuses data and makes decisions based on an inference process. As a result, commands are sent to actuators for changing system behavior, such as turning the air conditioner or lights on or off. Zyrianoff \textit{et al.} \cite{Borelli-2018},  Kamienski \textit{et al.} \cite{Kamienski-2015-context}, Kamienski \textit{et al.} \cite{Kamienski-2017-1}, and Pramudianto \textit{et al. }\cite{Pramudianto-2016} addressed similar situations.

This IoT application uses the following patterns:

\begin{itemize}
    \item Data Ingestion: Real Time Streaming Pattern for the system to capture and transfer data from sensors; Protocol Converter Pattern to convert different protocols used by different sensors
    \item Data Interaction: Publish/Subscribe Pattern to send sensor data to a high-capacity computing component
    \item Data Processing: Data Fusion and Data Analysis Pattern to process and manage context information
    \item Data Storage: SQL Pattern to store data in a relational model
    \item Data Visualization: RESTful API so the application can query information through an interaction component
     
    \item Connectors: Light Interaction to connect sensors to the MQTT broker; API to connect the cloud to query and management applications
\end{itemize}

\subsection{Precision Irrigation for Smart Agriculture}

Agricultural production plays a vital role in each nation's economy and has a continual improvement of its processes and techniques. However, agriculture consumes most of the freshwater available in the world. With climate change, the introduction of IoT-based technologies in the field is essential to secure our future through precision irrigation. Water usage can be substantially reduced, but fears of decrease in productivity due to water stress on plants lead farmers to over irrigate, which can lead to waste through the infiltration of water into the soil, as well as the energy used for irrigation. These also are a challenge for the sustainability of the planet. Then comes the need for greater water control in the irrigation of crops.

\begin{figure}[ht]
 \centering
 \includegraphics[scale=1.8]{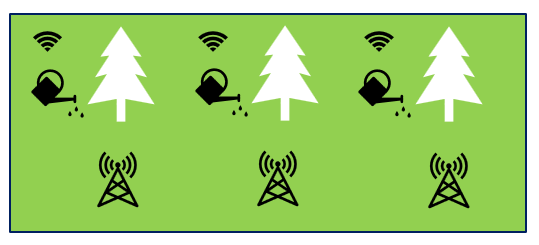}
 \caption{Smart irrigation. In this scenario, the goal is to automate irrigation according to the crops and climate of the region. }
 \label{fig:smart-irrigation}
\end{figure}

Precision agriculture collects and interprets vast amounts of data for enhanced field management. The precise management of irrigation plays a significant role in the continuous increase of production. Firstly, there is a need to identify the tools to acquire an enormous amount of data generated by sensors and other sources to be analyzed and compared. Some of the difficulties in adopting precision irrigation are related to data transferring, handling and processing, and the high cost of investing in hardware solutions to save this massive amount of data. Kamienski \textit{et al.} \cite{Kamienski-2019}, Liqiang \textit{et al.} \cite{Liqiang-2011}, Shahanas \textit{et al.} \cite{Shahanas-2016}, Ntuli \textit{et al.}\cite{Ntuli-2016}, Rad \textit{et al.}\cite{Rad-2015}, Robles \textit{et al.}\cite{Robles-2014}, Xiao \textit{et al.}\cite{Xiao-2010} presented possible solutions for agriculture irrigation. 

\begin{figure}[ht]
 \centering
 \includegraphics[scale=1.8]{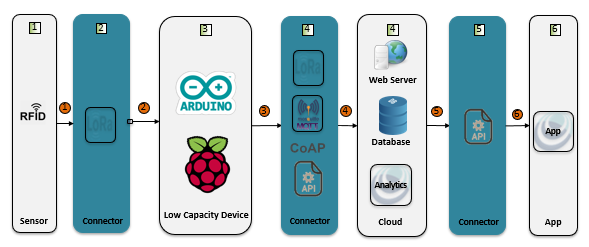}
 \caption{Reference architecture for irrigation management applications for smart agriculture. A soil moisture sensor sends  data to a low capacity processing component using LoRaWAN technology. The low capacity device sends data to a large capacity server that resides in the cloud, where databases, data distribution components, and irrigation model computing services are located. Data analysis techniques are used in these models to generate the irrigation plan.}
 \label{fig:arch-ref-irrigation}
\end{figure}

Figure \ref{fig:arch-ref-irrigation} shows this reference architecture: a device, typically a soil moisture sensor, sends messages to an interaction component, such as a LoraWAN gateway, residing in a low capacity computing element, such as a Raspberry Pi. From the gateway, data is sent by MQTT to the IoT platform and to components that implement models to estimate water needs and optimize irrigation, located in the cloud. As a result, an irrigation plan is sent to the irrigation system, which may involve controlling the actuators directly (such as valves, pumps, and sprinklers) or the interaction with a third-party system through an API.

This IoT application uses the following patterns:
\begin{itemize}
    \item	Data Ingestion: Real Time Streaming Pattern for the system to capture and transfer data from sensors; Protocol Converter Pattern to convert different protocols used by different sensors
    \item	Data Interaction: Publish/Subscribe Pattern to send sensor data to a high-capacity computing component
    \item	Data Processing: Data Fusion and Data Analysis Pattern to process and manage context information related to irrigation.
    \item	Data Storage: SQL Pattern to store data in a relational model. NoSQL pattern to store unstructured and semi-structured data. HDFS pattern to create a data lake and store the data to be processed.
    \item	Data Visualization: RESTful API so the application can query information through an interaction component
    \item	Connectors: Light Interaction to connect sensors to the MQTT broker or the LoRaWAN gateway; API to connect the cloud to query and management applications
\end{itemize}

\section{Discussion and Challenges}
\label{sec:discussion-challenges}

\subsection{Message exchanges in IoT}

The use of data interaction components and data ingestion components patterns ensures support to message delivery, mediation between different communication protocols, and message consumption tracking. Large companies want to build a solution that analyzes substantial amounts of data in a short time, but artificial intelligence algorithms need reliable data to work correctly \cite{Zhou-2017-ml, Tonyali-2018}. Obtaining data from different sources and reliably moving it is still a challenge, and the existing batch-based solutions have not solved the problem \cite{Mahdavinejad-2018, Wang-2016}.

These systems are extremely important when large amounts of data need to be moved between the most diverse components of different categories, according to section \ref{sec:padroes-aquiteturais-existentes}. From an IoT perspective, the use of data interaction components and data ingestion components patterns is necessary because the volume of messages grows exponentially with the increase of devices connected to the Internet. Few IoT solutions propose using these components to move data between components massively in IoT scenarios \cite{Mahdavinejad-2018, Zeng-2015-real-time}.

\subsection{IoT Device Interoperability}

Component connectors are essential for an IoT solution as they can negatively influence the performance of this ecosystem.  Zyrianoff \textit{et al.} \cite{Borelli-2018} carried out some performance tests on a smart building scenario, and the results show the influence of the use of data fusion components and connectors and data inference (data processing). An architecture that processes data in sequentially connected components that are connected sequentially is not always an adequate option because the message needs to arrive in one component to be transmitted to the next. This scheme generates processing bottlenecks throughout the component chain because if the message takes a while to be processed on a specific component, it delays the entire message exchange stream of the application. An alternative to a software architecture with sequentially connected components is to place a component that can centralize data sending and distribute to other components, such as FIWARE Orion. Thus, if any component delays the message processing, it only affects  a part of the message processing and sending stream, since the components are not connected in sequence.

Using standardized protocols for sending and receiving messages is a fair practice when choosing communication in a project. Although there are software components that convert protocols between the communication of two or more systems, using a single communication protocol facilitates the integration of projects that have multiple teams working on the same project. An example is FIWARE \cite{Fiware}, which facilitates the development of IoT solutions. All FIWARE components communicate via a single protocol, NGSI JSON,  which simplifies the development of the solution as a whole.

\subsection{IoT-sourced Data Storage}

With the decentralization of software architectures, new storage types, such as NoSQL, are emerging to enable scalability of IoT solutions. NoSQL DBMSs are used for specific problems. Hills \cite{Hills-2016} discusses a modeling approach that uses SQL and NoSQL to build software. New IoT platforms will use hybrid modeling to solve solutions to real scenarios in the future.

\subsection{Future works}

Future challenges include:

\begin{itemize}

    \item 	Understand the influence of creating hierarchies of low capacity communication and processing elements on a distributed architecture. Some IoT scenarios require the use of this type of knowledge, such as a scenario in which an ambulance is in an emergency and needs to pass through slow car traffic. The ambulance can communicate with edge (low capacity) devices that can make decisions even before sending data to software components located in the cloud.
    
    \item 	Study and propose an insight into the legal approach to data protection and how far applications can collect and process data from users. There are few discussions regarding the issue of data privacy, and the legislation is gradually adapting to the technological reality. It is necessary to understand how data is collected, analyzed, and stored by IoT devices. Privacy policies should explain the data lifecycle. The lack of these policies is a vulnerability to the data.
    
    \item 	Develop and research new IoT-specific code security solutions that can defend its systems against internal network attacks. IoT requires custom-made security solutions because, unlike a traditional computer, IoT elements usually: 1) have less processing capacity, memory and power supply; 2) execute tasks collaboratively; and 3) run eminently developed systems using C language or languages based on it. Since existing security proposals for computers that make up the traditional Internet do not take these characteristics into account, they are not always suitable for IoT.
    
    \item 	Propose new software architectures so that organizations can adapt legacy systems to the new near real-time data processing paradigm. For example, banking companies still use scripts that run during hours of low data processing and read files with financial records to insert data into storage components. This data movement could be treated as a stream and processed in near real time with Kappa architecture, for example.
    
    \item 	Propose new software architectures capable of ingesting data even with the emergence of new data formats and increasing data volume brought by IoT. Create new data ingestion components from new software architectures.

\end{itemize}

\section{Conclusion}
\label{sec:conclusion}

This paper presented an overview of software architecture patterns in an approach related to the development of IoT applications, as well as a classification of the major component classes that can be used in IoT software solutions. This component classification is expected to facilitate the development of IoT applications by  software developers, who can save hours to study to understand how to propose a software architecture for IoT. Besides, this paper discussed the main difficulties in creating a software architecture for IoT, which will help future developers to make better design decisions.

\section*{Acknowledgment}

The authors would like to thank to SWAMP Project.

\bibliographystyle{unsrt}  
\bibliography{references}  

\begin{thebibliography}{100}

\bibitem{Kruchten-2006}
PPhilippe {Kruchten}, Henk {Obbink}, and Judith {Stafford}.
\newblock The past, present, and future for software architecture.
\newblock {\em IEEE Software}, 23(2):22--30, March 2006.

\bibitem{shaw2006golden}
M.~{Shaw} and P.~{Clements}.
\newblock The golden age of software architecture.
\newblock {\em IEEE Software}, 23(2):31--39, March 2006.

\bibitem{Freeman-2004}
Elisabeth Freeman, Eric Freeman, Bert Bates, and Kathy Sierra.
\newblock {\em Head First Design Patterns}.
\newblock O' Reilly \& Associates, Inc., 2004.

\bibitem{Richards-2015}
Mark Richards.
\newblock {\em Software Architecture Patterns}.
\newblock O'Reilly Media, Inc., 2015.

\bibitem{Weyrich-2016}
Michael Weyrich and Christof Ebert.
\newblock Reference architectures for the internet of things.
\newblock {\em IEEE Software}, 33(1):112--116, Jan 2016.

\bibitem{Aksu-2018}
Hidayet Aksu, Leonardo Babun, Mauro Conti, Gabriele Tolomei, and Selcuk
  Uluagac.
\newblock Advertising in the iot era: Vision and challenges.
\newblock {\em IEEE Communications Magazine}, PP, 01 2018.

\bibitem{Taivalsaari-2017}
Antero {Taivalsaari} and Tommi {Mikkonen}.
\newblock A roadmap to the programmable world: Software challenges in the iot
  era.
\newblock {\em IEEE Software}, 34(1):72--80, Jan 2017.

\bibitem{Motta-2018}
Rebeca Motta, K\'{a}thia~M. de~Oliveira, and Guilherme~H. Travassos.
\newblock On challenges in engineering iot software systems.
\newblock In {\em Proceedings of the XXXII Brazilian Symposium on Software
  Engineering}, SBES '18, pages 42--51, New York, NY, USA, 2018. ACM.

\bibitem{Fowler-2002}
Martin Fowler.
\newblock {\em Patterns of Enterprise Application Architecture}.
\newblock Addison-Wesley Longman Publishing Co., Inc., Boston, MA, USA, 2002.

\bibitem{Leff-2001}
A.~{Leff} and J.~T. {Rayfield}.
\newblock Web-application development using the model/view/controller design
  pattern.
\newblock In {\em Proceedings Fifth IEEE International Enterprise Distributed
  Object Computing Conference}, pages 118--127, Sep. 2001.

\bibitem{Gamma-1994}
Erich Gamma, Richard Helm, Ralph Johnson, and John~M. Vlissides.
\newblock {\em Design Patterns: Elements of Reusable Object-Oriented Software}.
\newblock Addison-Wesley Professional, 1 edition, 1994.

\bibitem{Garlan-1993}
David Garlan and Mary Shaw.
\newblock An introduction to software architecture.
\newblock Technical report, Pittsburgh, PA, USA, 1994.

\bibitem{Heuzeroth-2003}
D.~{Heuzeroth}, T.~{Holl}, G.~{Hogstrom}, and W.~{Lowe}.
\newblock Automatic design pattern detection.
\newblock In {\em 11th IEEE International Workshop on Program Comprehension,
  2003.}, pages 94--103, May 2003.

\bibitem{Kruchten-2000}
Philippe Kruchten.
\newblock {\em The Rational Unified Process: An Introduction, Second Edition}.
\newblock 01 2000.

\bibitem{Bass-2012}
Bass Len, Clements Paul, , and Rick Kazman.
\newblock {\em Software Architecture In Practice}.
\newblock 01 2003.

\bibitem{ISO-42010}
Iso/iec/ieee systems and software engineering -- architecture description.
\newblock {\em ISO/IEC/IEEE 42010:2011(E) (Revision of ISO/IEC 42010:2007 and
  IEEE Std 1471-2000)}, pages 1--46, Dec 2011.

\bibitem{Bassi-2016}
Alessandro Bassi, Martin Bauer, Martin Fiedler, Thorsten Kramp, Rob van
  Kranenburg, Sebastian Lange, and Stefan Meissner.
\newblock {\em Enabling Things to Talk: Designing IoT Solutions with the IoT
  Architectural Reference Model}.
\newblock Springer Publishing Company, Incorporated, 1st edition, 2016.

\bibitem{Razzaque-2016}
Mohammad~Abdur Razzaque, Marija Milojevic-Jevric, Andrei Palade, and Siobhán
  Clarke.
\newblock Middleware for internet of things: a survey.
\newblock 3:1--1, 01 2015.

\bibitem{Baldauf-2007}
Matthias Baldauf, Schahram Dustdar, and Florian Rosenberg.
\newblock A survey on context\&\#45;aware systems.
\newblock {\em Int. J. Ad Hoc Ubiquitous Comput.}, 2(4):263--277, June 2007.

\bibitem{Moreno-2017}
M.~V. {Moreno}, F.~{Terroso-Sáenz}, A.~{González-Vidal}, M.~{Valdés-Vela},
  A.~F. {Skarmeta}, M.~A. {Zamora}, and V.~{Chang}.
\newblock Applicability of big data techniques to smart cities deployments.
\newblock {\em IEEE Transactions on Industrial Informatics}, 13(2):800--809,
  April 2017.

\bibitem{AlNuaimi-2015}
Nader~Mohamed Eiman Al~Nuaimi, Hind Al~Neyadi and Jameela Al-Jaroodi.
\newblock Applications of big data to smart cities.
\newblock {\em Journal of Internet Services and Applications}, 6, 08 2015.

\bibitem{Kamienski-2018-Tracker}
Carlos~A. Kamienski, Fabrizio~F. Borelli, Gabriela~O. Biondi, Isaac Pinheiro,
  Ivan~D. Zyrianoff, and Marc Jentsch.
\newblock Context design and tracking for iot-based energy management in smart
  cities.
\newblock {\em IEEE Internet of Things Journal}, 5(2):687--695, April 2018.

\bibitem{Lee-Law-2017}
Wen-Tin Lee and Po-Jen Law.
\newblock A case study in applying security design patterns for iot software
  system.
\newblock In {\em 2017 International Conference on Applied System Innovation
  (ICASI)}, pages 1162--1165, May 2017.

\bibitem{Qanbari-2016}
Soheil Qanbari, Samim Pezeshki, Rozita Raisi, Samira Mahdizadeh, Rabee
  Rahimzadeh, Negar Behinaein, Fada Mahmoudi, Shiva Ayoubzadeh, Parham Fazlali,
  Keyvan Roshani, Azalia Yaghini, Mozhdeh Amiri, Ashkan Farivarmoheb, Arash
  Zamani, and Schahram Dustdar.
\newblock Iot design patterns: Computational constructs to design, build and
  engineer edge applications.
\newblock In {\em 2016 IEEE First International Conference on
  Internet-of-Things Design and Implementation (IoTDI)}, pages 277--282, April
  2016.

\bibitem{Brambilla-2017}
Marco Brambilla, Eric Umuhoza, and Roberto Acerbis.
\newblock Model-driven development of user interfaces for iot systems via
  domain-specific components and patterns.
\newblock {\em Journal of Internet Services and Applications}, 8(1):14, Sep
  2017.

\bibitem{Reinfurt-2016}
Lukas Reinfurt, Uwe Breitenb\"{u}cher, Michael Falkenthal, Frank Leymann, and
  Andreas Riegg.
\newblock Internet of things patterns.
\newblock In {\em Proceedings of the 21st European Conference on Pattern
  Languages of Programs}, EuroPlop '16, pages 5:1--5:21, New York, NY, USA,
  2016. ACM.

\bibitem{Koster-2019}
Michael Koster.
\newblock Design patterns are reusable solutions to common problems, 2019.

\bibitem{Silva-2013}
Welington~M. da~Silva, Alexandre Alvaro, Gustavo Tomas, Ricardo Afonso, Kelvin
  Dias, and Vinicius Garcia.
\newblock Smart cities software architectures: A survey.
\newblock pages 1722--1727, 03 2013.

\bibitem{Yin-2015}
Chuantao Yin, Zhang Xiong, Hui Chen, Jingyuan Wang, Daven Cooper, and Bertrand
  David.
\newblock A literature survey on smart cities.
\newblock {\em Science China Information Sciences}, 58, 08 2015.

\bibitem{Ray-2016}
Partha~Pratim Ray.
\newblock A survey on internet of things architectures.
\newblock {\em Journal of King Saud University - Computer and Information
  Sciences}, 30(3):291 -- 319, 2018.

\bibitem{Mahmoud-2015}
Rwan Mahmoud, Tasneem Yousuf, Fadi Aloul, and Imran Zualkernan.
\newblock Internet of things (iot) security: Current status, challenges and
  prospective measures.
\newblock In {\em 2015 10th International Conference for Internet Technology
  and Secured Transactions (ICITST)}, pages 336--341, Dec 2015.

\bibitem{Santana-2017}
Eduardo Felipe~Zambom Santana, Ana~Paula Chaves, Marco~Aurelio Gerosa, Fabio
  Kon, and Dejan Milojicic.
\newblock Software platforms for smart cities: Concepts, requirements,
  challenges, and a unified reference architecture.
\newblock {\em ACM Computing Surveys}, 2017.

\bibitem{Atzori-2010}
Luigi Atzori, Antonio Iera, and Giacomo Morabito.
\newblock The internet of things: A survey.
\newblock {\em Computer Networks}, 54(15):2787 -- 2805, 2010.

\bibitem{Kamienski-2017-1}
Carlos~Alberto Kamienski, Mark Jentsch, Markus Eisenhauer, Juisse Kiljander,
  Enrico Ferrera, Eduardo Souto Walter~Andrade Peter~Rosengren, Peter~Thestrup,
  and Djamel Sadok.
\newblock Application development for the internet of things: A context-aware
  mixed criticality systems development platform.
\newblock {\em Computer Communications}, 104:1--16, 2017.

\bibitem{Kamienski-2019}
C.~{Kamienski}, J.~{Soininen}, M.~{Taumberger}, S.~{Fernandes}, A.~{Toscano},
  T.~S. {Cinotti}, R.~F. {Maia}, and A.~T. {Neto}.
\newblock Swamp: an iot-based smart water management platform for precision
  irrigation in agriculture.
\newblock In {\em 2018 Global Internet of Things Summit (GIoTS)}, pages 1--6,
  June 2018.

\bibitem{Borelli-2018}
Ivan Zyrianoff, Fabrizio~F. Borelli, Alexandre Heideker, Gabriela~O. Biondi,
  and Carlos Kamienski.
\newblock Scalability of iot-enabled context-aware management systems for smart
  cities.
\newblock In {\em IEEE Symposium on Computers and Communications (ISCC)}, 2018.

\bibitem{Hall-1997}
D.~L. {Hall} and J.~{Llinas}.
\newblock An introduction to multisensor data fusion.
\newblock {\em Proceedings of the IEEE}, 85(1):6--23, Jan 1997.

\bibitem{Qiao-2015}
Lin Qiao, Yinan Li, Sahil Takiar, Ziyang Liu, Narasimha Veeramreddy, Min Tu,
  Ying Dai, Issac Buenrostro, Kapil Surlaker, Shirshanka Das, and Chavdar
  Botev.
\newblock Gobblin: Unifying data ingestion for hadoop.
\newblock {\em Proc. VLDB Endow.}, 8(12):1764--1769, August 2015.

\bibitem{Sawant-2013}
Nitin Sawant and Himanshu Shah.
\newblock {\em Big Data Application Architecture Q \& A: A Problem-Solution
  Approach}.
\newblock Apress, Berkeley, CA, 2013.

\bibitem{Lengyel-2015}
László Lengyel, Péter Ekler, Tamás Ujj, Tamás Balogh, and Hassan Charaf.
\newblock Sensorhub: An iot driver framework for supporting sensor networks and
  data analysis.
\newblock {\em International Journal of Distributed Sensor Networks},
  11(7):454379, 2015.

\bibitem{Huang-2014}
Sheng Huang, Yaoliang Chen, Xiaoyan Chen, Kai Liu, Xiaomin Xu, Chen Wang, Kevin
  Brown, and Inge Halilovic.
\newblock The next generation operational data historian for iot based on
  informix.
\newblock In {\em Proceedings of the 2014 ACM SIGMOD International Conference
  on Management of Data}, SIGMOD '14, pages 169--176, New York, NY, USA, 2014.
  ACM.

\bibitem{Bashir-2016}
Muhammad~Rizwan Bashir and Asif~Qumer Gill.
\newblock Towards an iot big data analytics framework: Smart buildings systems.
\newblock In {\em 2016 IEEE 18th International Conference on High Performance
  Computing and Communications; IEEE 14th International Conference on Smart
  City; IEEE 2nd International Conference on Data Science and Systems
  (HPCC/SmartCity/DSS)}, pages 1325--1332, Dec 2016.

\bibitem{Marosi-2018}
Attila~Csaba Marosi, Attila Farkas, and Robert Lovas.
\newblock An adaptive cloud-based iot back-end architecture and its
  applications.
\newblock In {\em 2018 26th Euromicro International Conference on Parallel,
  Distributed and Network-based Processing (PDP)}, pages 513--520, March 2018.

\bibitem{Miloslavskaya-2016}
Natalia Miloslavskaya and Alexander Tolstoy.
\newblock Big data, fast data and data lake concepts.
\newblock {\em Procedia Computer Science}, 88:300 -- 305, 2016.
\newblock 7th Annual International Conference on Biologically Inspired
  Cognitive Architectures, BICA 2016, held July 16 to July 19, 2016 in New York
  City, NY, USA.

\bibitem{Verma-2017}
S.~{Verma}, Y.~{Kawamoto}, Z.~M. {Fadlullah}, H.~{Nishiyama}, and N.~{Kato}.
\newblock A survey on network methodologies for real-time analytics of massive
  iot data and open research issues.
\newblock {\em IEEE Communications Surveys Tutorials}, 19(3):1457--1477,
  thirdquarter 2017.

\bibitem{Cenni-2017}
Daniele Cenni, Paolo Nesi, Gianni Pantaleo, and Imad Zaza.
\newblock Twitter vigilance: A multi-user platform for cross-domain twitter
  data analytics, nlp and sentiment analysis.
\newblock In {\em 2017 IEEE SmartWorld, Ubiquitous Intelligence Computing,
  Advanced Trusted Computed, Scalable Computing Communications, Cloud Big Data
  Computing, Internet of People and Smart City Innovation
  (SmartWorld/SCALCOM/UIC/ATC/CBDCom/IOP/SCI)}, pages 1--8, Aug 2017.

\bibitem{Colmenares-2017}
Juan {Colmenares}, Reza {Dorrigiv}, and Daniel {Waddington}.
\newblock A single-node datastore for high-velocity multidimensional sensor
  data.
\newblock In {\em 2017 IEEE International Conference on Big Data (Big Data)},
  pages 445--452, Dec 2017.

\bibitem{Ta-Shma-2018}
Paula Ta-Shma, Adnan Akbar, Guy Gerson-Golan, Guy Hadash, Francois Carrez, and
  Klaus Moessner.
\newblock An ingestion and analytics architecture for iot applied to smart city
  use cases.
\newblock {\em IEEE Internet of Things Journal}, 5(2):765--774, April 2018.

\bibitem{Tanenbaum-2006}
Andrew Tanenbaum and Maarten van Steen.
\newblock {\em Distributed Systems: Principles and Paradigms (2Nd Edition)}.
\newblock Prentice-Hall, Inc., Upper Saddle River, NJ, USA, 2006.

\bibitem{Endrei-2004}
Mark Endrei, Jenny Ang, Ali Arsanjani, Sook Chua, Philippe Comte, Pål
  Krogdahl~Min Luo, and Tony Newling.
\newblock {\em Patterns: Service-oriented Architecture and Web Services}.
\newblock IBM Corp., Riverton, NJ, USA, 2004.

\bibitem{Alonso-2004}
Gustavo Alonso, Fabio Casati, Harumi Kuno, and Vijay Machiraju.
\newblock {\em Web Services: Concepts, Architectures and Applications
  (Data-Centric Systems and Applications)}.
\newblock Springer-Verlag Berlin Heidelberg, 2004.

\bibitem{Esposte-2017}
Arthur de~M~Del~Esposte, Fabio Kon, Fabio Costa, and Nelson Lago.
\newblock Interscity: A scalable microservice-based open source platform for
  smart cities.
\newblock {\em Proceedings of the 6th International Conference on Smart Cities
  and Green ICT Systems}, pages 35--46, 2017.

\bibitem{battle2008bridging}
Robert Battle and Edward Benson.
\newblock Bridging the semantic web and web 2.0 with representational state
  transfer (rest).
\newblock {\em Journal of Web Semantics}, 6(1):61 -- 69, 2008.
\newblock Semantic Web and Web 2.0.

\bibitem{Almeida-2013}
Ricardo~Aparecido Perez~de Almeida, Michael Blackstock, Rodger Lea, Roberto
  Calderon, Antonio~Francisco do~Prado, and Helio~Crestana Guardia.
\newblock Thing broker: A twitter for things.
\newblock In {\em Proceedings of the 2013 ACM Conference on Pervasive and
  Ubiquitous Computing Adjunct Publication}, UbiComp '13 Adjunct, pages
  1545--1554, New York, NY, USA, 2013. ACM.

\bibitem{RESTFul}
Leonard Richardson and Sam Ruby.
\newblock {\em Restful Web Services}.
\newblock O'Reilly, first edition, 2007.

\bibitem{Akbar-2017}
A.~{Akbar}, A.~{Khan}, F.~{Carrez}, and K.~{Moessner}.
\newblock Predictive analytics for complex iot data streams.
\newblock {\em IEEE Internet of Things Journal}, 4(5):1571--1582, Oct 2017.

\bibitem{nodered2019}
OpenJS Foundation.
\newblock Node-red: Flow-based programming for the internet of things, 2019.

\bibitem{oasis2018mqtt}
OASIS.
\newblock Oasis committee specification 02, 2019.

\bibitem{garg2013apache}
N~Garg.
\newblock Apache kafka, 2013.

\bibitem{Tarkoma-2012}
Sasu Tarkoma.
\newblock {\em Publish / Subscribe Systems: Design and Principles}.
\newblock Wiley Publishing, 1st edition, 2012.

\bibitem{Chen-2014}
Hsiang~Wen Chen and Fuchun~Joseph Lin.
\newblock Converging mqtt resources in etsi standards based m2m platform.
\newblock In {\em 2014 IEEE International Conference on Internet of Things
  (iThings), and IEEE Green Computing and Communications (GreenCom) and IEEE
  Cyber, Physical and Social Computing (CPSCom)}, pages 292--295, Sep. 2014.

\bibitem{Beighley-2007}
Lynn Beighley.
\newblock {\em Head First Sql}.
\newblock O'Reilly, first edition, 2007.

\bibitem{Rautmare-2016}
S.~{Rautmare} and D.~M. {Bhalerao}.
\newblock Mysql and nosql database comparison for iot application.
\newblock In {\em 2016 IEEE International Conference on Advances in Computer
  Applications (ICACA)}, pages 235--238, Oct 2016.

\bibitem{Phan-2014}
T.~A.~M. {Phan}, J.~K. {Nurminen}, and M.~{Di Francesco}.
\newblock Cloud databases for internet-of-things data.
\newblock In {\em 2014 IEEE International Conference on Internet of Things
  (iThings), and IEEE Green Computing and Communications (GreenCom) and IEEE
  Cyber, Physical and Social Computing (CPSCom)}, pages 117--124, Sep. 2014.

\bibitem{Tropashko-2007}
Vadim Tropashko and Donald Burleson.
\newblock {\em SQL Design Patterns: Expert Guide to SQL Programming}.
\newblock Rampant TechPress, 2007.

\bibitem{Lee-2015-2}
C.~{Lee} and Y.~{Zheng}.
\newblock Sql-to-nosql schema denormalization and migration: A study on content
  management systems.
\newblock In {\em 2015 IEEE International Conference on Systems, Man, and
  Cybernetics}, pages 2022--2026, Oct 2015.

\bibitem{Han-2011}
{Jing Han}, {Haihong E}, {Guan Le}, and {Jian Du}.
\newblock Survey on nosql database.
\newblock In {\em 2011 6th International Conference on Pervasive Computing and
  Applications}, pages 363--366, Oct 2011.

\bibitem{Stonebraker-2010}
Michael Stonebraker.
\newblock Sql databases v. nosql databases.
\newblock {\em Commun. ACM}, 53(4):10--11, April 2010.

\bibitem{Silberschatz-1996}
Avi Silberschatz, Henry Korth, and S~Sudarshan. ~.
\newblock Data models.
\newblock {\em ACM Comput. Surv.}, 28(1):105--108, March 1996.

\bibitem{Cecchinel-2014}
C.~{Cecchinel}, M.~{Jimenez}, S.~{Mosser}, and M.~{Riveill}.
\newblock An architecture to support the collection of big data in the internet
  of things.
\newblock In {\em 2014 IEEE World Congress on Services}, pages 442--449, June
  2014.

\bibitem{Grover-2015}
Mark Grover, Ted Malaska, Jonathan Seidman, and Gwen Shapira.
\newblock {\em Hadoop Application Architectures: Designing Real-World Big Data
  Applications}.
\newblock O'Reilly Media, 2015.

\bibitem{Bengfort-2016}
Benjamin Bengfort and Jenny Kim.
\newblock {\em Data Analytics with Hadoop: An Introduction for Data
  Scientists}.
\newblock O'Reilly Media, Inc., 1st edition, 2016.

\bibitem{Erraissi-2017}
Allae Erraissi, Abdessamad Belangour, and Abderrahim Tragha.
\newblock A comparative study of hadoop-based big data architectures.
\newblock {\em International Journal of Web Applications}, 9:129--137, 12 2017.

\bibitem{Miner-2012}
Donald Miner and Adam Shook.
\newblock {\em MapReduce Design Patterns: Building Effective Algorithms and
  Analytics for Hadoop and Other Systems}.
\newblock O'Reilly Media, Inc., 1st edition, 2012.

\bibitem{Liu-2017}
Xiufeng Liu, Alfred Heller, and {Per Sieverts} Nielsen.
\newblock Citiesdata: a smart city data management framework.
\newblock {\em Knowledge and Information Systems}, 2017.

\bibitem{Bellini-2018}
Pierfrancesco Bellini, Paolo Nesi, Michela Paolucci, and Imad Zaza.
\newblock Smart city architecture for data ingestion and analytics: Processes
  and solutions.
\newblock In {\em 2018 IEEE Fourth International Conference on Big Data
  Computing Service and Applications (BigDataService)}, pages 137--144, March
  2018.

\bibitem{Dean-2008}
Jeffrey Dean and Sanjay Ghemawat.
\newblock Mapreduce: Simplified data processing on large clusters.
\newblock {\em Commun. ACM}, 51(1):107--113, January 2008.

\bibitem{Cormen-2009}
Thomas Cormen and Charles Leisersonand Ronald Rivestand~Clifford Stein.
\newblock {\em Introduction to Algorithms, Third Edition}.
\newblock The MIT Press, 3rd edition, 2009.

\bibitem{Wu-2017-big-data}
Dongyao Wu, Sherif Sakr, and Liming Zhu.
\newblock {\em Big Data Programming Models}, pages 31--63.
\newblock Springer International Publishing, Cham, 2017.

\bibitem{Belcastro-2018}
Loris Belcastro, Fabrizio Marozzo, and Domenico Talia.
\newblock Programming models and systems for big data analysis.
\newblock {\em International Journal of Parallel, Emergent and Distributed
  Systems}, 34(6):632--652, 2019.

\bibitem{Hadoop}
Apache hadoop, 2019.

\bibitem{Qureshi-2019}
Nawab Muhammad~Faseeh Qureshi, Isma~Farah Siddiqui, Mukhtiar~Ali Unar,
  Muhammad~Aslam Uqaili, Choon~Sung Nam, Dong~Ryeol Shin, Jaehyoun Kim,
  Ali~Kashif Bashir, and Asad Abbas.
\newblock An aggregate mapreduce data block placement strategy for wireless iot
  edge nodes in smart grid.
\newblock {\em Wireless Personal Communications}, 106(4):2225--2236, Jun 2019.

\bibitem{Hughes-89}
John Hughes.
\newblock Why functional programming matters.
\newblock {\em Comput. J.}, 32(2):98--107, April 1989.

\bibitem{Hudak-1989}
Paul Hudak.
\newblock Conception, evolution, and application of functional programming
  languages.
\newblock {\em ACM Comput. Surv.}, 21(3):359--411, September 1989.

\bibitem{Spark}
Apache spark, 2019.

\bibitem{Flink}
Apache flink, 2019.

\bibitem{Feng-2018}
Feng Ye, Peng Zhang, Cheng Hu, Songjie Zhu, and Ling Li.
\newblock The tentative research of hydrological iot data processing system
  based on apache flink.
\newblock In {\em Service-Oriented Computing -- ICSOC 2018 Workshops}, pages
  161--168, Cham, 2019. Springer International Publishing.

\bibitem{Wiley-2015}
Joshua Wiley and Larry Pace.
\newblock {\em Beginning R: An Introduction to Statistical Programming}.
\newblock Apress, Berkely, CA, USA, 2nd edition, 2015.

\bibitem{Nesa-2018}
N.~{Nesa}, T.~{Ghosh}, and I.~{Banerjee}.
\newblock Outlier detection in sensed data using statistical learning models
  for iot.
\newblock In {\em 2018 IEEE Wireless Communications and Networking Conference
  (WCNC)}, pages 1--6, April 2018.

\bibitem{Grover-2015-sql-like}
A.~{Grover}, J.~{Gholap}, V.~P. {Janeja}, Y.~{Yesha}, R.~{Chintalapati},
  H.~{Marwaha}, and K.~{Modi}.
\newblock Sql-like big data environments: Case study in clinical trial
  analytics.
\newblock In {\em 2015 IEEE International Conference on Big Data (Big Data)},
  pages 2680--2689, Oct 2015.

\bibitem{Klein-1999}
Lawrence Klein.
\newblock {\em Sensor and Data Fusion Concepts and Applications}.
\newblock Society of Photo-Optical Instrumentation Engineers (SPIE),
  Bellingham, WA, USA, 2nd edition, 1999.

\bibitem{Llinas-1989}
J.~{Llinas}, D.~L. {Hall}, and E.~{Waltz}.
\newblock Data fusion technology forecast for c/sup 3/mis.
\newblock In {\em 1989 Third International Conference on Command, Control,
  Communications and Management Information Systems}, pages 148--158, May 1989.

\bibitem{Kamienski-2015-context}
Carlos Kamienski, Fabrizio Borelli, Gabriela Biondi, Wiliam Rosa, Isaac
  Pinheiro, Ivan Zyrianoff, Djamel Sadok, and Ferry Pramudianto.
\newblock Context-aware energy efficiency management for smart buildings.
\newblock In {\em 2015 IEEE 2nd World Forum on Internet of Things (WF-IoT)},
  pages 699--704, Dec 2015.

\bibitem{Nagl-2006}
C.~{Nagl}, F.~{Rosenberg}, and S.~{Dustdar}.
\newblock Vidre--a distributed service-oriented business rule engine based on
  ruleml.
\newblock In {\em 2006 10th IEEE International Enterprise Distributed Object
  Computing Conference (EDOC'06)}, pages 35--44, Oct 2006.

\bibitem{Abiteboul-2008}
Serge Abiteboul, Ohad Greenshpan, and Tova Milo.
\newblock Modeling the mashup space.
\newblock In {\em Proceedings of the 10th ACM Workshop on Web Information and
  Data Management}, WIDM '08, pages 87--94, New York, NY, USA, 2008. ACM.

\bibitem{Blackstock-2012}
Michael Blackstock and Rodger Lea.
\newblock Iot mashups with the wotkit.
\newblock pages 159--166, 10 2012.

\bibitem{Soukaras-2015}
Dimitris Soukaras, Pankesh Patel, Hui Song, and Sanjay Chaudhary.
\newblock Iotsuite: a toolsuite for prototyping internet of things
  applications.
\newblock {\em The 4th International Workshop on Computing and Networking for
  Internet of Things (ComNet-IoT), co-located with 16th International
  Conference on Distributed Computing and Networking (ICDCN)}, 2015.

\bibitem{Merlino-2014}
G.~{Merlino}, D.~{Bruneo}, S.~{Distefano}, F.~{Longo}, and A.~{Puliafito}.
\newblock Stack4things: Integrating iot with openstack in a smart city context.
\newblock In {\em 2014 International Conference on Smart Computing Workshops},
  pages 21--28, Nov 2014.

\bibitem{Conti-2018}
Mauro Conti, Ali Dehghantanha, Katrin Franke, and Steve Watson.
\newblock Internet of things security and forensics: Challenges and
  opportunities.
\newblock {\em Future Generation Computer Systems}, 78:544 -- 546, 2018.

\bibitem{Andrea-2015}
Ioannis Andrea, Chrysostomos Chrysostomou, and George Hadjichristofi.
\newblock Internet of things: Security vulnerabilities and challenges.
\newblock In {\em 2015 IEEE Symposium on Computers and Communication (ISCC)},
  pages 180--187, July 2015.

\bibitem{Zhou-2017-security}
J.~{Zhou}, Z.~{Cao}, X.~{Dong}, and A.~V. {Vasilakos}.
\newblock Security and privacy for cloud-based iot: Challenges.
\newblock {\em IEEE Communications Magazine}, 55(1):26--33, January 2017.

\bibitem{Choudhury-2017}
T.~{Choudhury}, A.~{Gupta}, S.~{Pradhan}, P.~{Kumar}, and Y.~S. {Rathore}.
\newblock Privacy and security of cloud-based internet of things (iot).
\newblock In {\em 2017 3rd International Conference on Computational
  Intelligence and Networks (CINE)}, pages 40--45, Oct 2017.

\bibitem{Hafiz-2007}
M.~{Hafiz}, P.~{Adamczyk}, and R.~E. {Johnson}.
\newblock Organizing security patterns.
\newblock {\em IEEE Software}, 24(4):52--60, July 2007.

\bibitem{Alaba-2017}
Fadele~Ayotunde Alaba, Mazliza Othman, Ibrahim Abaker~Targio Hashem, and Faiz
  Alotaibi.
\newblock Internet of things security: A survey.
\newblock In {\em 2018 International Conference on Advanced Science and
  Engineering (ICOASE)}, pages 162--166, Oct 2018.

\bibitem{Gubbi-2013}
Jayavardhana Gubbi, Rajkumar Buyya, Slaven Marusic, and Marimuthu Palaniswami.
\newblock Internet of things (iot): A vision, architectural elements, and
  future directions.
\newblock {\em Future Generation Computer}, 29(7):1645--1660, 2013.

\bibitem{Gaur-2015}
Aditya Gaur, Bryan Scotney, Gerard Parr, and Sally McClean.
\newblock Smart city architecture and its applications based on iot.
\newblock {\em Procedia Computer Science}, 52:1089 -- 1094, 2015.
\newblock The 6th International Conference on Ambient Systems, Networks and
  Technologies (ANT-2015), the 5th International Conference on Sustainable
  Energy Information Technology (SEIT-2015).

\bibitem{Nehme-2010}
R.~V. {Nehme}, H.~{Lim}, and E.~{Bertino}.
\newblock Fence: Continuous access control enforcement in dynamic data stream
  environments.
\newblock In {\em 2010 IEEE 26th International Conference on Data Engineering
  (ICDE 2010)}, pages 940--943, March 2010.

\bibitem{Botta-2016}
Alessio Botta, Walter de~Donato, Valerio Persico, and Antonio Pescapé.
\newblock Integration of cloud computing and internet of things: A survey.
\newblock {\em Future Generation Computer Systems}, 56:684 -- 700, 2016.

\bibitem{Eschenauer-2002}
Laurent Eschenauer and Virgil Gligor.
\newblock A key-management scheme for distributed sensor networks.
\newblock In {\em Proceedings of the 9th ACM Conference on Computer and
  Communications Security}, CCS '02, pages 41--47, New York, NY, USA, 2002.
  ACM.

\bibitem{Wachter-2018}
Sandra Wachter.
\newblock Normative challenges of identification in the internet of things:
  Privacy, profiling, discrimination, and the {GDPR}.
\newblock {\em Computer Law {\&} Security Review}, 34(3):436--449, jun 2018.

\bibitem{Lu-2017}
R.~{Lu}, K.~{Heung}, A.~H. {Lashkari}, and A.~A. {Ghorbani}.
\newblock A lightweight privacy-preserving data aggregation scheme for fog
  computing-enhanced iot.
\newblock {\em IEEE Access}, 5:3302--3312, 2017.

\bibitem{Bormann-2019}
Carsten Bormann, Mehmet Ersue, and Ari Keränen.
\newblock {Terminology for Constrained-Node Networks}.
\newblock RFC 7228, May 2014.

\bibitem{Farrell-2018}
U.~{Raza}, P.~{Kulkarni}, and M.~{Sooriyabandara}.
\newblock Low power wide area networks: An overview.
\newblock {\em IEEE Communications Surveys Tutorials}, 19(2):855--873,
  Secondquarter 2017.

\bibitem{Nakamura-2007}
Eduardo Nakamura, Antonio Loureiro, and Alejandro Frery.
\newblock Information fusion for wireless sensor networks: Methods, models, and
  classifications.
\newblock {\em ACM Computing Surveys (CSUR)}, 39:9, 09 2007.

\bibitem{Castanedo-2013}
Federico Castanedo.
\newblock A review of data fusion techniques.
\newblock 2013, 10 2013.

\bibitem{Klein-1993}
Lawrence~A. Klein.
\newblock {\em Sensor and Data Fusion Concepts and Applications}.
\newblock Society of Photo-Optical Instrumentation Engineers (SPIE),
  Bellingham, WA, USA, 1993.

\bibitem{Esper}
Esper, 2019.

\bibitem{Drools}
Jboss drools, 2019.

\bibitem{OpenRules}
Openrules, 2019.

\bibitem{ThingsBoard}
Thingsboard 2.0, 2019.

\bibitem{Pramudianto-2016}
Ferry Pramudianto, Markus Eisenhauer, Carlos~Alberto Kamienski, Djamel Sadok,
  and Eduardo~J. Souto.
\newblock Connecting the internet of things rapidly through a model driven
  approach.
\newblock In {\em 3rd {IEEE} World Forum on Internet of Things, WF-IoT 2016,
  Reston, VA, USA, December 12-14, 2016}, pages 135--140, 2016.

\bibitem{Henderson-2006}
Cal Henderson.
\newblock {\em Building Scalable Web Sites: Building, Scaling, and Optimizing
  the Next Generation of Web Applications}.
\newblock O'Reilly Media, Inc., 2006.

\bibitem{Hiromoto-2017}
R.~E. {Hiromoto}, M.~{Haney}, and A.~{Vakanski}.
\newblock A secure architecture for iot with supply chain risk management.
\newblock In {\em 2017 9th IEEE International Conference on Intelligent Data
  Acquisition and Advanced Computing Systems: Technology and Applications
  (IDAACS)}, volume~1, pages 431--435, Sep. 2017.

\bibitem{Kaisler-2013}
S.~{Kaisler}, F.~{Armour}, J.~A. {Espinosa}, and W.~{Money}.
\newblock Big data: Issues and challenges moving forward.
\newblock In {\em 2013 46th Hawaii International Conference on System
  Sciences}, pages 995--1004, Jan 2013.

\bibitem{Marjani-2017}
M.~{Marjani}, F.~{Nasaruddin}, A.~{Gani}, A.~{Karim}, I.~A.~T. {Hashem},
  A.~{Siddiqa}, and I.~{Yaqoob}.
\newblock Big iot data analytics: Architecture, opportunities, and open
  research challenges.
\newblock {\em IEEE Access}, 5:5247--5261, 2017.

\bibitem{Kafka}
Apache kafka, 2019.

\bibitem{Nifi}
Apache nifi, 2019.

\bibitem{Flume}
Apache flume, 2019.

\bibitem{Shreedharan-2014}
Hari Shreedharan.
\newblock {\em Using Flume: Flexible, Scalable, and Reliable Data Streaming}.
\newblock O'Reilly Media, 2014.

\bibitem{GraphQL}
Graphql. a query language for your api, 2019.

\bibitem{Ultralight}
Ultralight 2.0, 2019.

\bibitem{LoRaWAN}
Lorawan, 2019.

\bibitem{Fiware-IoT-Agent}
Fiware-iot-agent, 2019.

\bibitem{Cai-2017}
H.~Cai, B.~Xu, L.~Jiang, and A.~V. Vasilakos.
\newblock Iot-based big data storage systems in cloud computing: Perspectives
  and challenges.
\newblock {\em IEEE Internet of Things Journal}, 4(1):75--87, Feb 2017.

\bibitem{Cai-2014}
Hongming Cai, Li~Da Xu, Cheng Xie, Shaojun Qin, and Lihong Jiang.
\newblock Iot-based configurable information service platform for product
  lifecycle management.
\newblock {\em IEEE Transactions on Industrial Informatics}, 10(2):1558--1567,
  May 2014.

\bibitem{Cure-2012}
Olivier Curé, Fadhela Kerdjoudj, David Faye, Chan~Le Duc, and Myriam Lamolle.
\newblock On the potential integration of an ontology-based data access
  approach in nosql stores.
\newblock In {\em 2012 Third International Conference on Emerging Intelligent
  Data and Web Technologies}, pages 166--173, Sep. 2012.

\bibitem{Zhu-2013-storage}
Minbo Li, Zhu Zhu, and Guangyu Chen.
\newblock A scalable and high-efficiency discovery service using a new storage.
\newblock In {\em 2013 IEEE 37th Annual Computer Software and Applications
  Conference}, pages 754--759, July 2013.

\bibitem{Ma-2012}
Youzhong Ma, Jia Rao, Weisong Hu, Xiaofeng Meng, Xu~Han, Yu~Zhang, Yunpeng
  Chai, and Chunqiu Liu.
\newblock An efficient index for massive iot data in cloud environment.
\newblock In {\em Proceedings of the 21st ACM International Conference on
  Information and Knowledge Management}, CIKM '12, pages 2129--2133, New York,
  NY, USA, 2012. ACM.

\bibitem{Mallapuram-2017}
S.~Mallapuram, N.~Ngwum, F.~Yuan, C.~Lu, and W.~Yu.
\newblock Smart city: The state of the art, datasets, and evaluation platforms.
\newblock In {\em 2017 IEEE/ACIS 16th International Conference on Computer and
  Information Science (ICIS)}, pages 447--452, May 2017.

\bibitem{Jiang-2014}
Lihong Jiang, Li~Da Xu, Hongming Cai, Zuhai Jiang, Fenglin Bu, and Boyi Xu.
\newblock An iot-oriented data storage framework in cloud computing platform.
\newblock {\em IEEE Transactions on Industrial Informatics}, 10(2):1443--1451,
  May 2014.

\bibitem{Fiware}
Fiware, 2019.

\bibitem{Fiware-Orion}
Fiware orion, 2019.

\bibitem{Fiware-Cosmos}
Fiware cosmos, 2019.

\bibitem{Quantum-Leap}
Quantum leap, 2019.

\bibitem{CrateDB}
Cratedb, 2019.

\bibitem{Ji-2019}
W.~{Ji}, J.~{Xu}, H.~{Qiao}, M.~{Zhou}, and B.~{Liang}.
\newblock Visual iot: Enabling internet of things visualization in smart
  cities.
\newblock {\em IEEE Network}, 33(2):102--110, March 2019.

\bibitem{Kimball-2004}
Ralph Kimball and Joe Caserta.
\newblock {\em The Data Warehouse ETL Toolkit: Practical Techniques for
  Extracting, Cleaning, Conforming and Delivering Data}.
\newblock John Wiley \&\#38; Sons, Inc., USA, 2004.

\bibitem{Bansal-2015}
Srividya~K. Bansal and Sebastian Kagemann.
\newblock Integrating big data: A semantic extract-transform-load framework.
\newblock {\em Computer}, 48(3):42--50, Mar 2015.

\bibitem{Vassiliadis-2009}
Panos Vassiliadis.
\newblock A survey of extract-transform-load technology.
\newblock {\em International Journal of Data Warehousing and Mining}, 5:1--27,
  07 2009.

\bibitem{Fang-2015}
H.~{Fang}.
\newblock Managing data lakes in big data era: What's a data lake and why has
  it became popular in data management ecosystem.
\newblock In {\em 2015 IEEE International Conference on Cyber Technology in
  Automation, Control, and Intelligent Systems (CYBER)}, pages 820--824, June
  2015.

\bibitem{Marz-2013}
Nathan Marz.
\newblock {\em Big data: principles and best practices of scalable realtime
  data systems}.
\newblock O'Reilly Media, [S.l.], 2013.

\bibitem{Kiran-2015}
Mariam Kiran, Peter Murphy, Inder Monga, Jon Dugan, and Sartaj~Singh Baveja.
\newblock Lambda architecture for cost-effective batch and speed big data
  processing.
\newblock In {\em 2015 IEEE International Conference on Big Data (Big Data)},
  pages 2785--2792, Oct 2015.

\bibitem{Kreps-2014}
Questioning the lambda architecture, 2014.

\bibitem{Lin-2017}
Jie Lin, Wei Yu, Nan Zhang, Xinyu Yang, Hanlin Zhang, and Wei Zhao.
\newblock A survey on internet of things: Architecture, enabling technologies,
  security and privacy, and applications.
\newblock {\em IEEE Internet of Things Journal}, 4(5):1125--1142, Oct 2017.

\bibitem{Bishop-2006}
Christopher~M. Bishop.
\newblock {\em Pattern Recognition and Machine Learning (Information Science
  and Statistics)}.
\newblock Springer-Verlag, Berlin, Heidelberg, 2006.

\bibitem{Golden-2001}
R.M. Golden.
\newblock Statistical pattern recognition.
\newblock In Neil~J. Smelser and Paul~B. Baltes, editors, {\em International
  Encyclopedia of the Social \& Behavioral Sciences}, pages 15040 -- 15044.
  Pergamon, Oxford, 2001.

\bibitem{Jain-2000}
Anil~K. Jain, Robert Duin, and Jianchang Mao.
\newblock Statistical pattern recognition: A review.
\newblock {\em IEEE Trans. Pattern Anal. Mach. Intell.}, 22:4--37, 01 2000.

\bibitem{Mahdavinejad-2018}
Mohammad~Saeid Mahdavinejad, Mohammadreza Rezvan, Mohammadamin Barekatain,
  Peyman Adibi, Payam Barnaghi, and Amit~P. Sheth.
\newblock Machine learning for internet of things data analysis: a survey.
\newblock {\em Digital Communications and Networks}, 4(3):161 -- 175, 2018.

\bibitem{Roffia-2018}
Luca Roffia, Paolo Azzoni, Cristiano Aguzzi, Fabio Viola, Francesco Antoniazzi,
  and Tullio Cinotti.
\newblock Dynamic linked data: A sparql event processing architecture.
\newblock {\em Future Internet}, 10:36, 04 2018.

\bibitem{Chen-2015}
Feng Chen, Pan Deng, Jiafu Wan, Daqiang Zhang, Athanasios~V. Vasilakos, and
  Xiaohui Rong.
\newblock Data mining for the internet of things: Literature review and
  challenges.
\newblock {\em International Journal of Distributed Sensor Networks},
  11(8):431047, 2015.

\bibitem{Tsai-2014}
C.~{Tsai}, C.~{Lai}, M.~{Chiang}, and L.~T. {Yang}.
\newblock Data mining for internet of things: A survey.
\newblock {\em IEEE Communications Surveys Tutorials}, 16(1):77--97, First
  2014.

\bibitem{Li-2018}
H.~{Li}, K.~{Ota}, and M.~{Dong}.
\newblock Learning iot in edge: Deep learning for the internet of things with
  edge computing.
\newblock {\em IEEE Network}, 32(1):96--101, Jan 2018.

\bibitem{Khanna-2016}
Abhirup Khanna and Rishi Anand.
\newblock Iot based smart parking system.
\newblock In {\em 2016 International Conference on Internet of Things and
  Applications (IOTA)}, pages 266--270, Jan 2016.

\bibitem{Pham-2015}
T.~N. {Pham}, M.~{Tsai}, D.~B. {Nguyen}, C.~{Dow}, and D.~{Deng}.
\newblock A cloud-based smart-parking system based on internet-of-things
  technologies.
\newblock {\em IEEE Access}, 3:1581--1591, 2015.

\bibitem{Kalluri-2018}
Balaji Kalluri, Clayton Miller, Bharath Seshadri, and Arno Schlueter.
\newblock A cyber-physical middleware platform for buildings in smart cities.
\newblock pages 645--652, 10 2018.

\bibitem{Minoli-2017}
D.~{Minoli}, K.~{Sohraby}, and B.~{Occhiogrosso}.
\newblock Iot considerations, requirements, and architectures for smart
  buildings—energy optimization and next-generation building management
  systems.
\newblock {\em IEEE Internet of Things Journal}, 4(1):269--283, Feb 2017.

\bibitem{Liqiang-2011}
Zhao Liqiang, Yin Shouyi, Liu Leibo, Zhang Zhen, and Wei Shaojun.
\newblock A crop monitoring system based on wireless sensor network.
\newblock {\em Procedia Environmental Sciences}, 11:558–565, 12 2011.

\bibitem{Shahanas-2016}
K.~Mohammed Shahanas and P.~Bagavathi Sivakumar.
\newblock Framework for a smart water management system in the context of smart
  city initiatives in india.
\newblock {\em Procedia Computer Science}, 92:142 -- 147, 2016.
\newblock 2nd International Conference on Intelligent Computing, Communication
  \& Convergence, ICCC 2016, 24-25 January 2016, Bhubaneswar, Odisha, India.

\bibitem{Ntuli-2016}
Nonhlanhla Ntuli and Adnan Abu-Mahfouz.
\newblock A simple security architecture for smart water management system.
\newblock {\em Procedia Computer Science}, 83:1164--1169, 04 2016.

\bibitem{Rad-2015}
Ciprian-Radu Rad, Olimpiu Hancu, Ioana-Alexandra Takacs, and Gheorghe Olteanu.
\newblock Smart monitoring of potato crop: A cyber-physical system architecture
  model in the field of precision agriculture.
\newblock {\em Agriculture and Agricultural Science Procedia}, 6:73--79, 12
  2015.

\bibitem{Robles-2014}
Tom\'{a}s Robles, Ram\'{o}n Alcarria, Diego Mart\'{\i}n, Augusto Morales,
  Mariano Navarro, Rodrigo Calero, Sofia Iglesias, and Manuel L\'{o}pez.
\newblock An internet of things-based model for smart water management.
\newblock In {\em Proceedings of the 2014 28th International Conference on
  Advanced Information Networking and Applications Workshops}, WAINA '14, pages
  821--826, Washington, DC, USA, 2014. IEEE Computer Society.

\bibitem{Xiao-2010}
K.H. Xiao, D.Q. Xiao, and X.W. Luo.
\newblock Smart water-saving irrigation system in precision agriculture based
  on wireless sensor network.
\newblock {\em Trans. Chin. Soc. Agric. Eng.}, 26:170--175, 11 2010.

\bibitem{Zhou-2017-ml}
Lina Zhou, Shimei Pan, Jianwu Wang, and Athanasios~V. Vasilakos.
\newblock Machine learning on big data: Opportunities and challenges.
\newblock {\em Neurocomputing}, 237:350 -- 361, 2017.

\bibitem{Tonyali-2018}
Samet Tonyali, Kemal Akkaya, Nico Saputro, A.~Selcuk Uluagac, and Mehrdad
  Nojoumian.
\newblock Privacy-preserving protocols for secure and reliable data aggregation
  in iot-enabled smart metering systems.
\newblock {\em Future Generation Computer Systems}, 78:547 -- 557, 2018.

\bibitem{Wang-2016}
Hai Wang, Zeshui Xu, Hamido Fujita, and Shousheng Liu.
\newblock Towards felicitous decision making: An overview on challenges and
  trends of big data.
\newblock {\em Information Sciences}, 367-368:747 -- 765, 2016.

\bibitem{Zeng-2015-real-time}
J.~Liu Z.~Zheng, P.~Wang and S.~Sun.
\newblock Real-time big data processing framework: Challenges and solutions.
\newblock {\em Applied Mathematics and Information Sciences}, 9:3169--3190, 01
  2015.

\bibitem{Hills-2016}
Ted Hills.
\newblock {\em NoSQL and SQL Data Modeling: Bringing Together Data, Semantics,
  and Software}.
\newblock Technics Publications, New York, NY, USA, 2016.

\end{thebibliography}


\end{document}